\documentclass[prd,twocolumn,showpacs,superscriptaddress,preprintnumbers,nofootinbib,
amsmath,amssymb]{revtex4-1}

\usepackage{graphicx} 
\usepackage{tikz}
\usetikzlibrary{arrows}

\usepackage[colorlinks=true,urlcolor=blue,linkcolor=blue,citecolor=blue]{hyperref}
\usepackage[normalem]{ulem}
\usepackage[capitalize]{cleveref}
 
\usepackage{comment}
\usepackage{flushend}

\usepackage[T1]{fontenc} 

\usepackage{mathtools}
\usepackage{xcolor}
\usepackage{graphicx}
\usepackage{dcolumn}
\usepackage{bm}
\usepackage{multirow}
\usepackage{physics}
\usepackage{color, soul}
\usepackage{epstopdf}
\usepackage{float}
\usepackage{subcaption} 

\newcommand{\Beq}{\begin{equation}\begin{aligned}}
\newcommand{\Eeq}{\end{aligned}\end{equation}}

\begin{document}


\title{Little Red Dots from Small-Scale Primordial Black Hole Clustering} 

\author{Borui Zhang}
\affiliation{Department of Physics, Tsinghua University, Beijing 100084, China}

\author{Wei-Xiang Feng}
\email{wxfeng@mail.tsinghua.edu.cn}
\affiliation{Department of Physics, Tsinghua University, Beijing 100084, China}

\author{Haipeng An}
\email{anhp@mail.tsinghua.edu.cn}
\affiliation{Department of Physics, Tsinghua University, Beijing 100084, China}
\affiliation{Center for High Energy Physics, Tsinghua University, Beijing 100084, China}


\begin{abstract}
The James Webb Space Telescope (JWST) observations have identified a class of compact galaxies at high redshifts (\(4 \lesssim z \lesssim 8\)), dubbed ``little red dots'' (LRDs). The supermassive black holes (SMBHs) of $10^{5\text{--}8}{\rm\,M}_{\odot}$ in LRDs favor a heavy-seed origin. We propose a mechanism for their formation: Clusters of primordial black holes, formed through long-short mode coupling on small scales in the early Universe, undergo sequential mergers over extended timescales. This mechanism can evade cosmic microwave background distortions and result in heavy-seed SMBHs via runaway mergers. We employ Monte Carlo simulations to solve the Smoluchowski coagulation equation and determine the runaway merging timescale. The resulting stochastic gravitational wave background offers a distinctive signature of this process, and our model suggests highly spinning SMBHs at their formation if they inherit the spin residual of the cluster from tidal fields.
\end{abstract}

\maketitle

\paragraph*{Introduction and Summary.---}
The James Webb Space Telescope (JWST) has recently discovered a class of compact, red-hued galaxies at high redshifts (\(4\lesssim z\lesssim 8\)), dubbed ``little red dots'' (LRDs)\;\cite{Matthee:2023utn,Pacucci:2023oci,Gentile:2024uiw,akins2024cosmos,williams2024galaxies}, for which the origin remains open. Moreover, around 80\% of these galaxies display broad Balmer emission lines, suggesting that they host active galactic nuclei (AGN) with central supermassive black holes (SMBHs) of $10^{5\textup{--}8}{\rm\,M}_{\odot}$\;\cite{Harikane:2023aa,Maiolino:2023bpi,kocevski2023hidden,kokorev_uncover_2023,killi2024deciphering,kokorev_census_2024,wang_rubies_2024,Durodola:2024bom,Ananna:2024jug,kocevski_rise_2025,Furtak_2024,Maiolino:2025tih}. 

While extreme stellar densities could be an alternative of broad-line LRDs to AGN, these stellar configurations are probably short-lived\,\cite{Baggen:2024vef,Guia:2024toq,Carranza-Escudero:2025ved,Pacucci:2025ojp}. SMBHs in LRDs could be primordial black holes (PBHs) from the high-density perturbations in the early Universe. Nevertheless, PBHs produced from direct collapse are excluded for heavy seeds $\gtrsim10^4{\rm\,M}_\odot$ by cosmic microwave background (CMB) distortions given the nearly Gaussian primordial density perturbations\;\cite{Sasaki:2018dmp,Carr:2020gox}. Even if the constraint is evaded, they generally harbor small spin at formation\;\cite{Mirbabayi:2019uph,DeLuca:2019buf}.

The story changes if PBHs were, however, clustered right after their formation. Clustering of PBHs can occur through various mechanisms, such as local-type non-Gaussianity\;\cite{Byrnes:2011ri,Byrnes:2012yx,Young:2013oia,Young:2014oea,Young:2015kda,Tada:2015noa,Franciolini:2018vbk,Desjacques:2018wuu,Ali-Haimoud:2018dau,Young:2019gfc,Suyama:2019cst,Atal:2020igj,DeLuca:2022bjs,DeLuca:2022uvz}, closed domain walls\;\cite{Khlopov:2004sc,Dokuchaev:2004kr,Belotsky:2018wph}, multistream inflation\;\cite{Ding:2019tjk,Huang:2023mwy}, long-range scalar forces\;\cite{Amendola:2017xhl,Savastano:2019zpr,Flores:2020drq}, quantum diffusion\;\cite{Ezquiaga:2019ftu,Ezquiaga:2022qpw,Animali:2024jiz}, or correlated bubble collisions\;\cite{DeLuca:2021mlh}. However, it has been recently shown that PBH clustering is inevitable even without an exotic mechanism, but when the extreme-value statistics is properly considered for rare events of PBH formation\;\cite{Choi:2025eqn}.

In this \emph{Letter}, we investigate whether a high-redshift spinning SMBH can originate from successive mergers of PBHs in an initial small-scale cluster, which avoids the CMB constraints. Crucially, we compute the associated stochastic gravitational wave (GW) background of this process with a reasonable comoving number density of clusters across cosmic evolution that can explain the SMBHs in LRDs. In addition, we speculate that highly spinning SMBHs may inherit the spin residual of PBH clusters from tidal interactions. These features are distinct from massive PBHs of direct collapse\;\cite{Hooper:2023nnl,Qin:2025ymc,Guo:2026cuv,Allegrini:2026jqt}.

Single PBHs virialized in a cluster with number density $n_{\rm cl}$ will form binaries via the emission of gravitational radiation during close encounters and eventually merge. The fraction of PBHs that will coalesce over a time period $\Delta t$ can be computed as:
\begin{equation}
\frac{\Delta n_{\rm cl}}{n_{\rm cl}}=\frac{1}{2}\sigma_{\rm merg}\,v_{\rm rel}\,n_{\rm cl}\,\Delta t\,,
\end{equation}
where
\begin{equation}
\sigma_{\rm merg}
\!=2\pi\!\left(\frac{85\pi}{6\sqrt{2}}\right)^{\!\!2/7}\!\!\frac{G^2(m_1\!+\!m_2)^{10/7}m_1^{2/7}m_2^{2/7}}{c^{10/7}v_{\rm rel}^{18/7}}
\label{eq:merg_sigma}
\end{equation}
is the merging cross section\;\cite{Quinlan:1989,Mouri:2002mc,Feng:2024obn}.
Here, $G$ is the Newton's constant, $c$ is the speed of light, $m_1$ and $m_2$ are the masses of the two PBHs in a close encounter, and their relative velocity is $v_{\rm rel}\approx v_{\rm vir}=\sqrt{GM_{\rm cl}/r_{\rm vir}}$, the virial velocity of the cluster, where $M_{\rm cl}$ is the cluster mass within the virial radius $r_{\rm vir}$. To form a SMBH via mergers of an initial population of $N_{\rm cl}$ PBHs of equal mass $m_{\rm pbh}=30{\rm\,M}_\odot$, we can estimate the timescale over which half of the PBHs coalesce into larger ones, by
\begin{align}
&t_{\rm merg}\sim\frac{1}{n_{\rm cl}\,\sigma_{\rm merg}\,v_{\rm vir}}\simeq1.07{\rm\,Gyr}\left(\frac{N_{\rm cl}}{10^5}\right)^{-1}\nonumber\\
&\times\left(\frac{m_{\rm pbh}}{30{\rm\,M}_\odot}\right)^{\!-2}\!\!\left(\frac{M_{\rm cl}}{3\!\times\!10^6{\rm\,M}_\odot}\right)^{\!11/14}\!\!\left(\frac{r_{\rm vir}}{0.02\rm\,pc}\right)^{\!31/14}\!\!\!,
\end{align}
which indicates that $n_{\rm cl}=3N_{\rm cl}/4\pi r_{\rm vir}^3\gtrsim3\times10^9{\rm\,pc^{-3}}$ is required for the emergence of a $\sim10^6{\rm\,M}_\odot$ SMBH within the first billion years after the Big Bang. 
This estimate is conservative, as during mergers the merging cross section increases as PBHs coalesce into bigger ones, until the \emph{runaway merger} results. Then the question is: how can such a dense cluster be generated in the early Universe?

As a proof of concept, clusters of this kind may arise from the modulation of the overdensity field at the PBH formation scale ($k_s$) by long-wavelength ($k_l$) modes on larger scales if the primordial power spectrum peaks at these two modes. Here we consider $k_s\simeq 2.5{\rm\,pc^{-1}}$ at the horizon-exit for PBHs of mass $m_{\rm pbh}\simeq30{\rm\,M}_\odot$\;\cite{Atal:2020igj}. The long-mode spatial modulation is given by\;\cite{Atal:2020igj}:
\begin{align}\label{long_short_coupling}
\nu(\mathbf{x})=\nu_{g}\left(1+\eta\,\Phi_{l}(\mathbf{x})\right)\,,
\end{align}
where \(\eta\) is the coupling strength between long and short modes, \(\Phi_l(\mathbf{x})\) is the long-mode field, and \(\nu_g \equiv \delta_c / \bar{\sigma}_{s}\) is the reduced threshold of the short mode without modulation, with \(\delta_c \simeq 0.414\) during the radiation era\;\cite{Harada:2013epa}, and \(\bar{\sigma}_{s}^2 = \langle\delta_s^2\rangle\) the variance of \(\delta_s\) averaged over the whole space. Both the long- and short-mode density fields are assumed to follow Gaussian distributions. In this study, we fix $\nu_g=8.5$ to generate a sufficiently large initial clustering amplitude and to achieve an adequate PBH abundance (see Eqs.\,\ref{eq:pbh_abundance1} and \ref{eq:pbh_abundance2}), resulting in $\bar{\sigma}_s\simeq0.0487$. The variance of the long-mode, $\sigma_l^2$, is set to be $\sigma_l\simeq0.0064$ in order to evade the CMB $\mu$-distortion constraint\;\cite{Sasaki:2018dmp,Carr:2020gox}.

\begin{figure}[H]
	\centering
	\includegraphics[width=0.439\textwidth]{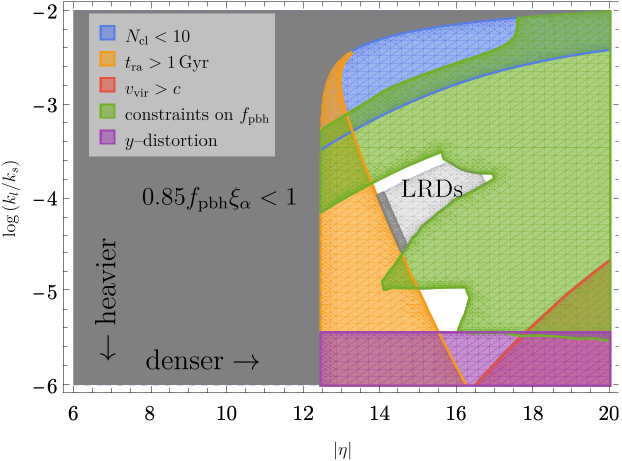}
	\caption{
    The $|\eta|\textup{--}(k_l/k_s)$ parameter plane for $\nu_g=8.5$, $\kappa\equiv\sigma_{l}/\bar{\sigma}_{s}=0.1313$, chosen to satisfy the stringent CMB $\mu$-distortion constraint\;\cite{Chluba:2012we,Chluba:2012gq,Kawasaki:2012kn,Kohri:2014lza,Nakama:2017xvq}, and the heavy-gray region is below the PBH cluster formation threshold. The blue is where the clusters are too light ($M_{\rm cl}=N_{\rm cl}m_{\rm pbh}\lesssim300{\rm\,M}_\odot$) to form SMBHs. The yellow is where the clusters are too sparse to have runaway mergers in $1{\rm\,Gyr}$; while the clusters in the red region are sufficiently dense, with virial velocity $v_{\rm vir}>c$, to have direct collapse. The green is excluded from observations on PBH abundance $f_{\rm pbh}$\;\cite{Carr:2016drx,Sasaki:2018dmp,Carr:2020gox}. The \(y\)-distortion constraint (purple) excludes clusters of \(M_{\rm cl}\gtrsim10^{11}{\rm\,M}_{\odot}\) that would result in ultra-massive black holes. The light-gray denotes the parameter space that permits the emergence of SMBHs ($10^{5\textup{--}8}{\rm\,M}_{\odot}$) in LRDs at redshifts $z\gtrsim5$, where the darker band exhibits the runaway merger time $t_{\rm ra}\gtrsim400{\rm\,Myr}$.}
	\label{fig:constraint}
 \end{figure}

Fig.\,\ref{fig:constraint} summarizes the parameter space in terms of the modulation strength $|\eta|$ and the comoving wavenumber ratio $k_l/k_s$, which determine the cluster's number density $n_{\rm cl}$ and mass $M_{\rm cl}$, respectively. Qualitatively, a stronger $|\eta|$ results in a denser cluster; while a smaller $k_l/k_s$ gives a heavier cluster. As will be discussed in the following sections, the light-gray region denotes where PBH clusters can form and explain SMBHs in LRDs via runaway mergers, consistent with observational constraints.

\paragraph*{Initial Small-Scale PBH Clustering.---} 

According to Eq.\,\ref{long_short_coupling}, for \( |\eta| \neq 0 \), the long-mode-induced modulation can generate a significant two-point correlation that remains nearly constant within the clustering scale \( r_{\text{cl}}\sim 1/k_l \):
\begin{align}
\xi_{\rm pbh}(r)\simeq\left\{ \begin{matrix}\xi_\alpha, & r\lesssim r_{\rm cl}; \\
0, &r> r_{\rm cl}\,, \end{matrix}\right.
\end{align}
where the subscript denotes coupling strength $\alpha(\eta, \kappa)\equiv \delta_c\eta\kappa$ with $\kappa\equiv\sigma_{l}/\bar{\sigma}_{s}\simeq0.1313$ the ratio of variance of long to short mode.
In our setting, $\xi_\alpha\sim10^6$ given $|\eta|\sim14$.
The fraction of PBHs relative to the total dark matter abundance,
defined as \(f_{\rm pbh}\equiv\Omega_{\rm pbh}/\Omega_{\rm dm}\), also depends on the long-short mode coupling \( |\eta| \), and is given by\;\cite{Sasaki:2018dmp,Carr:2020gox}:
\begin{equation}
f_{\rm pbh}\simeq  2.8\times10^{7}\left(\frac{g_\star}{106.75}\right)^{-1/4}\!\left(\frac{30{\rm\,M}_\odot}{m_{\rm pbh}}\right)^{1/2}\!\!\!\beta(m_{\rm pbh}),
\label{eq:pbh_abundance1}
\end{equation}
where $g_\star$ is the number of relativistic degrees of freedom at the time of PBH formation, and\;\cite{Atal:2020igj}
\begin{align}
\beta(m_{\rm pbh}) \approx 
0.1\,\text{erfc} \left( \frac{\nu_g}{\sqrt{2}} \frac{1}{\sqrt{1+\alpha^2}} \right)\,,
\label{eq:pbh_abundance2}
\end{align}
so the coupling strength enhances the PBH abundance. 

\begin{figure*}[t]
\centering
   \includegraphics[width=0.3296\textwidth]{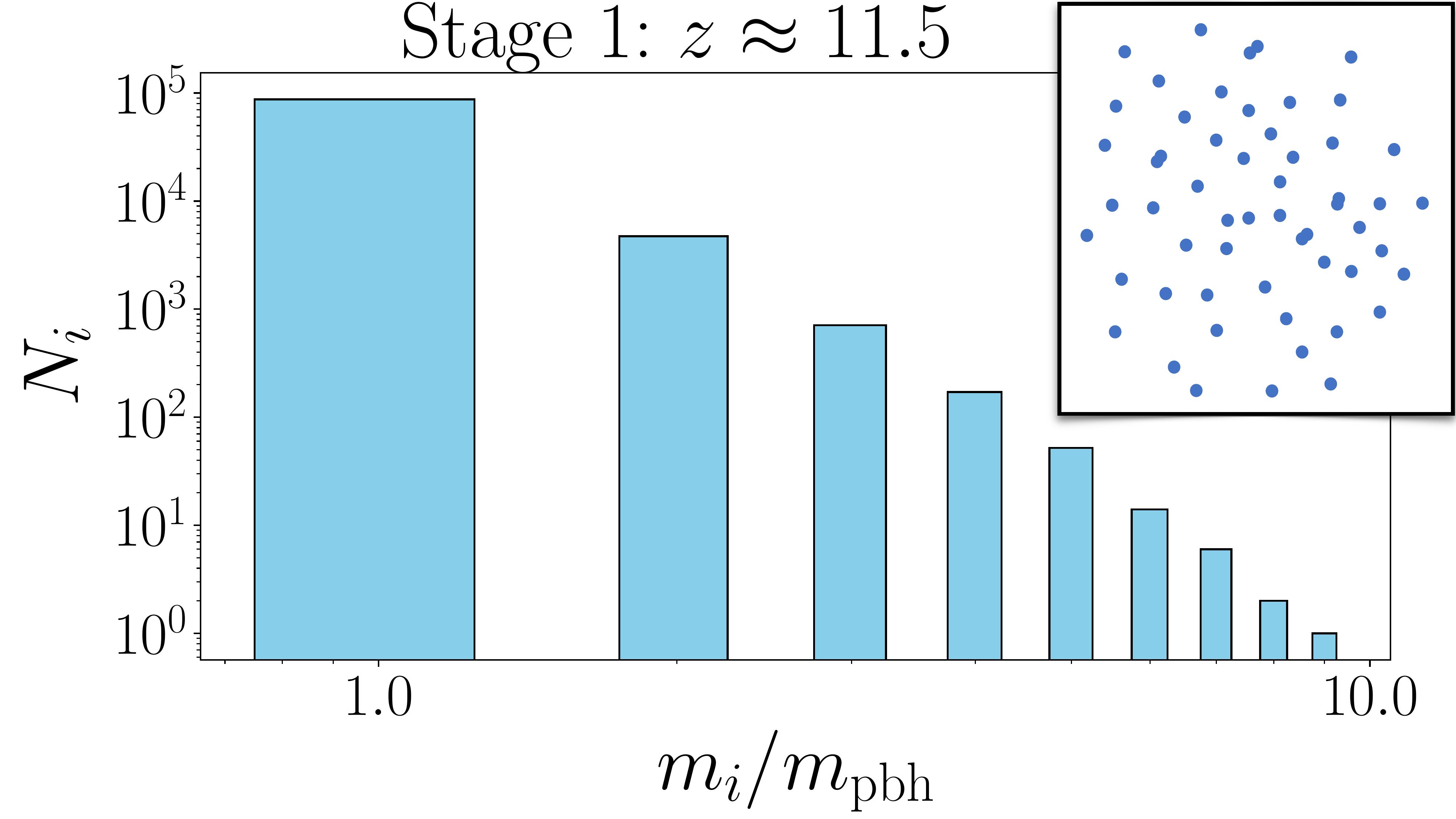}
   \includegraphics[width=0.3296\textwidth]{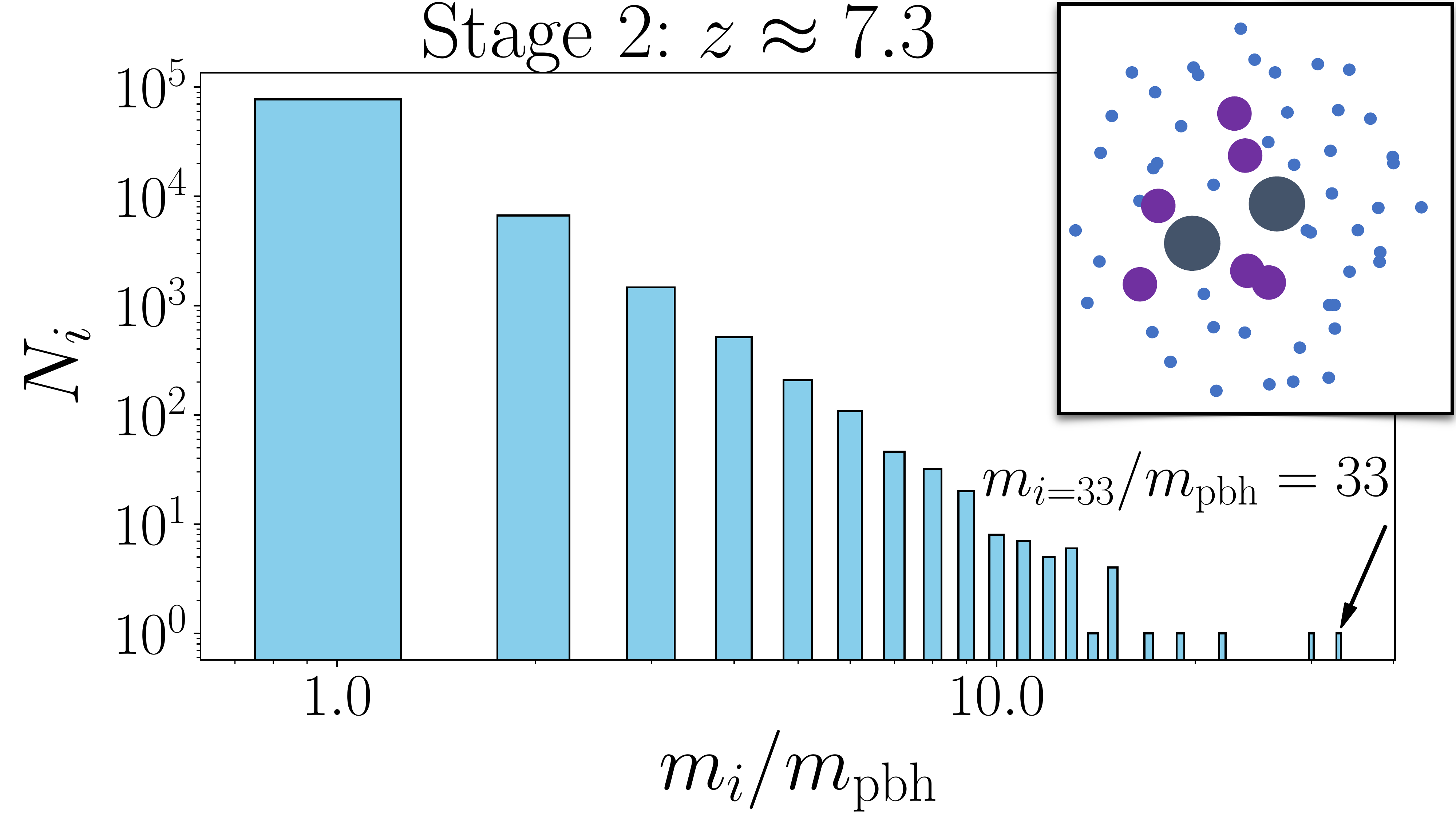}
   \includegraphics[width=0.3296\textwidth]{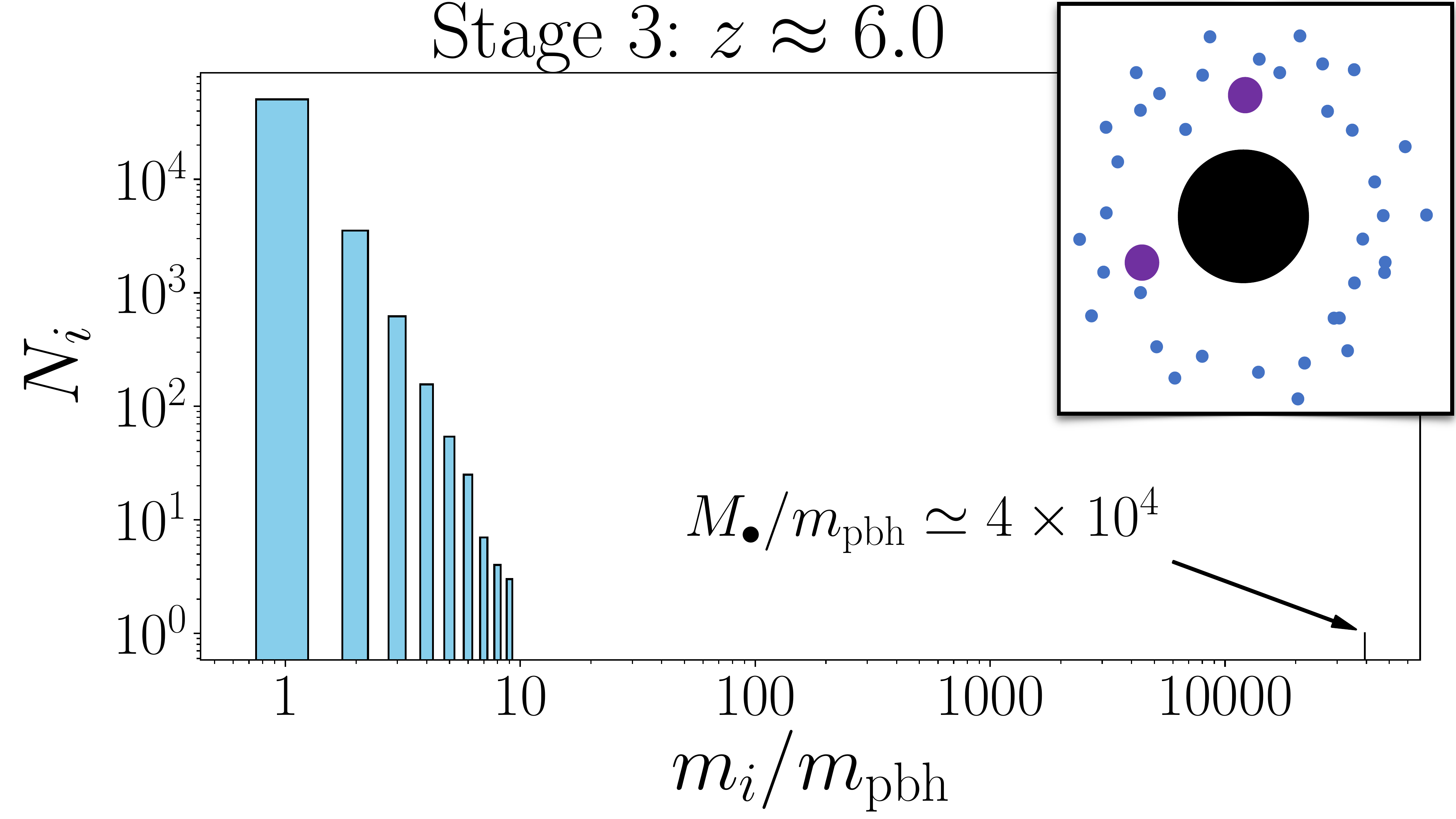}
    \caption{PBH mass population evolution in a cluster of $N_{\rm cl}=10^5$ from redshift $z\simeq11.5\textup{--}6.0$ with $n_{\rm cl}=2.0\times10^8{\rm\,pc^{-3}}$ and $v_{\rm vir}=512{\rm\,km\,s^{-1}}$ corresponding to $|\eta|\simeq14.34$ and $\log\,(k_l/k_s)\simeq-4.38$ in Fig.\,\ref{fig:constraint}.}
   \label{fig:combined}
\end{figure*}
The properties of the PBH cluster depend on the combination $f_{\rm pbh}\xi_\alpha$, which is a function of $\alpha(\eta, \kappa)$ and $k_l/k_s$. In particular, the condition $0.85f_{\rm pbh}\xi_\alpha>1$ (as shown in Fig.\,\ref{fig:constraint}) must be satisfied for decoupling before the matter-radiation equality. It turns out $f_{\rm pbh}\sim10^{-5}$ is sufficient to explain the SMBH abundance given the strong correlation $\xi_\alpha\sim10^6$. The cluster parameters are summarized below\;\cite{DeLuca:2022bjs,DeLuca:2022uvz} (see also Supplemental Material).

\paragraph*{The Number Density.} After the cluster decouples from the Universe's expansion, its number density can be treated as fixed\;\cite{DeLuca:2022bjs,DeLuca:2022uvz}:
\begin{align}
n_{\text{cl}} \simeq2\times10^8{\rm\,pc^{-3}}\left(\frac{f_{\rm pbh}\xi_\alpha}{25}\right)^4\left(\frac{C}{20}\right)\left(\frac{30{\rm\,M}_{\odot}}{m_{\text{pbh}}}\right),
\end{align}
where the compactness parameter $C=20$\;\cite{Kolb:1994fi} is adopted to characterize the cluster's overdensity after virialization. The number density \( n_{\text{cl}} \) is only determined by the combination \( f_{\rm pbh} \xi_\alpha \), and therefore remains constant after decoupling regardless of the cluster size $r_{\rm cl} \sim 1/k_l$ at formation. A strong coupling \( \eta \) boosts the clustering via $n_{\rm cl}\propto(f_{\rm pbh}\xi_\alpha)^4$. 

\paragraph*{The Cluster Mass.}
The total mass of the cluster is given by:
\begin{align}
M_{\text{cl}} \simeq 3\times10^6{\rm\,M}_{\odot}\left(\frac{f_{\rm pbh} \xi_\alpha}{25}\right)\left(\frac{r_{\rm cl}}{10\,\text{kpc}}\right)^3\,,
\end{align}
which scales as \( r_{\rm cl}^3 \sim 1/k_l^3 \), implying that the wavelength of the long mode primarily determines the total mass of the cluster and ultimately sets the mass scale of the final SMBH. As \( n_{\rm cl} \) is nearly independent of $r_{\rm cl}$, a longer-wavelength mode (smaller $k_l$) results in a larger cluster volume and hence a greater number of PBHs, $N_{\rm cl}=M_{\rm cl}/m_{\rm pbh}$. Therefore, for a fixed $M_{\rm cl}$, the number of PBHs in the cluster is inversely proportional to their individual masses $m_{\rm pbh}$.

\paragraph*{The Virial Radius and Velocity.}
The virial radius of the cluster after decoupling is computed as $r_{\rm vir}= \left( 3 M_{\text{cl}}/4 \pi m_{\rm pbh}n_{\text{cl}} \right)^{1/3}\propto\left(f_{\rm pbh}\xi_\alpha\right)^{-1}C^{-1/3}r_{\rm cl}$, which decreases with the coupling $|\eta|$. The virial velocity of the cluster is then given by $v_{\rm vir} = \sqrt{G M_{\text{cl}}/r_{\rm vir}}\propto\left(f_{\rm pbh}\xi_\alpha\right)C^{1/6}r_{\rm cl}$, which increases with the coupling $|\eta|$. As both $r_{\rm vir}$ and $v_{\rm vir}\propto r_{\rm cl}\sim 1/k_l$, the smaller the ratio $k_l / k_s$, the larger the $r_{\rm vir}$ and $v_{\rm vir}$. Given the above setting and $k_l / k_s\sim10^{-4}$, we will show that a cluster with  $r_{\rm vir}\sim0.05{\rm\,pc}$ and $v_{\rm vir}\sim500{\rm\,km\,s^{-1}}$ can explain the SMBHs in LRDs through runaway mergers.

\paragraph*{Runaway Mergers and the Gravitational Wave.---}
In the PBH-merger scenario, clusters at formation must not undergo direct collapse into an SMBH immediately after decoupling from the Hubble flow, such that \(r_{\rm vir} > 2GM_{\text{cl}}/c^2\) or \(v_{\rm vir} < c\) based on the hoop conjecture\;\cite{Flanagan:1991aa}.
For a virialized homogeneous cluster with gravitational potential energy $U=-0.6\,GM_{\rm cl}^2/r_{\rm vir}$, PBHs generally reach stable equilibrium with \(r_{\rm vir} < -0.335\,GM_{\rm cl}^2/E\), where the total energy is \(E=U/2\). Thus, according to the Antonov instability criterion\;\cite{Antonov:1977rw,Lynden-Bell:1968awc,Padmanabhan:1989vt,Chavanis:2001hd}, the system remains stable against gravothermal catastrophe until an SMBH emerges through successive mergers of PBHs, although the two processes may occur concurrently during the runaway evolution.

The initial cluster decouples with number density $n_{\rm cl}$ and PBHs of monochromatic mass $m_{\rm pbh}$. As PBH mergers are predominantly governed by two-body interactions, and binaries coalesce via the emission of gravitational radiation, we employ the \emph{Smoluchowski coagulation equation}\;\cite{Smoluchowski:1916,Smoluchowski:1917,Chandrasekhar:1943ws} to study the population evolution and the timescale leading to central massive black holes:
\begin{equation}
\frac{\rm d}{{\rm d}t}n_i=\frac{1}{2}\sum_{j+k=i}n_j n_k \mathcal{K}_{jk}
-n_i\sum_{j=1}^{N_{\rm cl}-i}n_j\mathcal{K}_{ij}\,,
\label{eq:Smoluchowski}
\end{equation}
where $n_i$ is the number density of PBH with mass $m_i=im_{\rm pbh}\,(i=1,2,...,N_{\rm cl})$. 
The first sum accounts for the gain in $n_i$ due to mergers of PBHs with $m_j+m_k=m_i$, while the second represents the loss in $n_i$ due to the mergers of PBHs of mass $m_i$ with any other PBH mass. 
The merger kernel $\mathcal{K}_{ij}=\langle\sigma_{\rm merg}(i,j)\,v_{\rm rel}\rangle$, capturing the rate of PBH mergers, is the velocity-weighted average of Eq.\,\ref{eq:merg_sigma}. Assuming the relative velocities of PBHs in the virialized cluster follow a Maxwellian distribution with energy equipartition, the kernel is given by\;\cite{Mouri:2002mc}
\begin{align}
\!\!\mathcal{K}_{ij}=\!\mathcal{A}\frac{G^2 m_{\rm pbh}^2}{c^3}\!\left( \frac{v_0}{c} \right)^{\!-11/7} 
\!\!\left( \frac{m_i m_j}{m_{\rm pbh}^2} \right)^{p}\!\!\left( \frac{m_i + m_j}{m_{\rm pbh}} \right)^{q}
\label{eq:merg_kernel}
\end{align}
with $\mathcal{A}=85^{2/7}(2\pi)^{11/14}3^{1/2}\Gamma(5/7)$, $v_0=\sqrt{3/5}\,v_{\rm vir}$ the initial (root-mean-square) velocity of PBHs, and $(p,q)=(15/14,9/14)$ with homogeneous spatial distribution. 

Fig.\,\ref{fig:combined} shows the PBH mass population evolution in an initial cluster of $N_{\rm cl}=10^5$ monochromatic PBHs from redshift $z\simeq11.5$ to $6.0$ with $n_{\rm cl}=2\times10^8{\rm\,pc^{-3}}$ and $v_{\rm vir}=512{\rm\,km\,s^{-1}}$, where we utilize the Monte Carlo scheme\;\cite{Zhang:2026vjk,Gillespie:1972,Gillespie:1975,garcia_monte_1987,liffman_direct_1992,babovsky_monte_1999,patterson_stochastic_2011,kotalczyk_monte_2017,tran2023fragmentation} to solve Eq.\,\ref{eq:Smoluchowski} using a full-conditioning method for sampling\;\cite{Gillespie:1975}. A SMBH of mass $M_\bullet\simeq10^6{\rm\,M}_\odot$ emerges by $z\sim6$ if the initial PBHs are of $m_{\rm pbh}=30{\rm\,M}_\odot$, which can mainly explain the LRDs that host $M_\bullet\simeq10^{5\textup{--}8}{\rm\,M}_{\odot}$, ranging from $z\simeq4\textup{--}8$, with or without the accretion of surrounding gas. In the simulation, the runaway merging timescale $t_{\rm ra}\approx0.2/n_{\rm cl}\mathcal{K}_{00}\lesssim910{\rm\,Myr}$ for completing the merger is smaller if $n_{\rm cl}$ is larger\,\cite{Zhang:2026vjk}. This is consistent with the given upper bound, $1/n_{\rm cl}\mathcal{K}_{00}$, in Refs.\,\cite{Mouri:2002mc,Zhang:2026vjk}. 

Since the coalescence time of two PBHs is much shorter than the disruption timescale caused by a third PBH in our typical cluster, even when the SMBH merges with smaller PBHs, the simulation is justified. In addition, energy equipartition, encoded in the merger kernel Eq.\,\ref{eq:merg_kernel}, is valid up until the onset of runaway mergers as the mass-segregation instability is triggered, while Eq.\,\ref{eq:merg_kernel} is still applicable even after the SMBH emergence as the relative velocity is mainly determined by the smaller PBHs.

The merger of a PBH cluster can be divided into three stages. The first stage involves only small-mass PBH mergers, initially at high redshifts. In the second stage, PBHs with intermediate masses begin to form and boost the merger process due to the increased merger cross section. The third stage eventually arises when the number density of these intermediate-mass PBHs grows beyond a certain threshold, and an SMBH will emerge with the surrounding small-mass PBHs, forming an \emph{extreme mass ratio inspiral system}. The overall merger processes will provide a distinct GW signature of this mechanism.

\begin{figure}[H]
	\centering
	\includegraphics[width=0.5\textwidth]{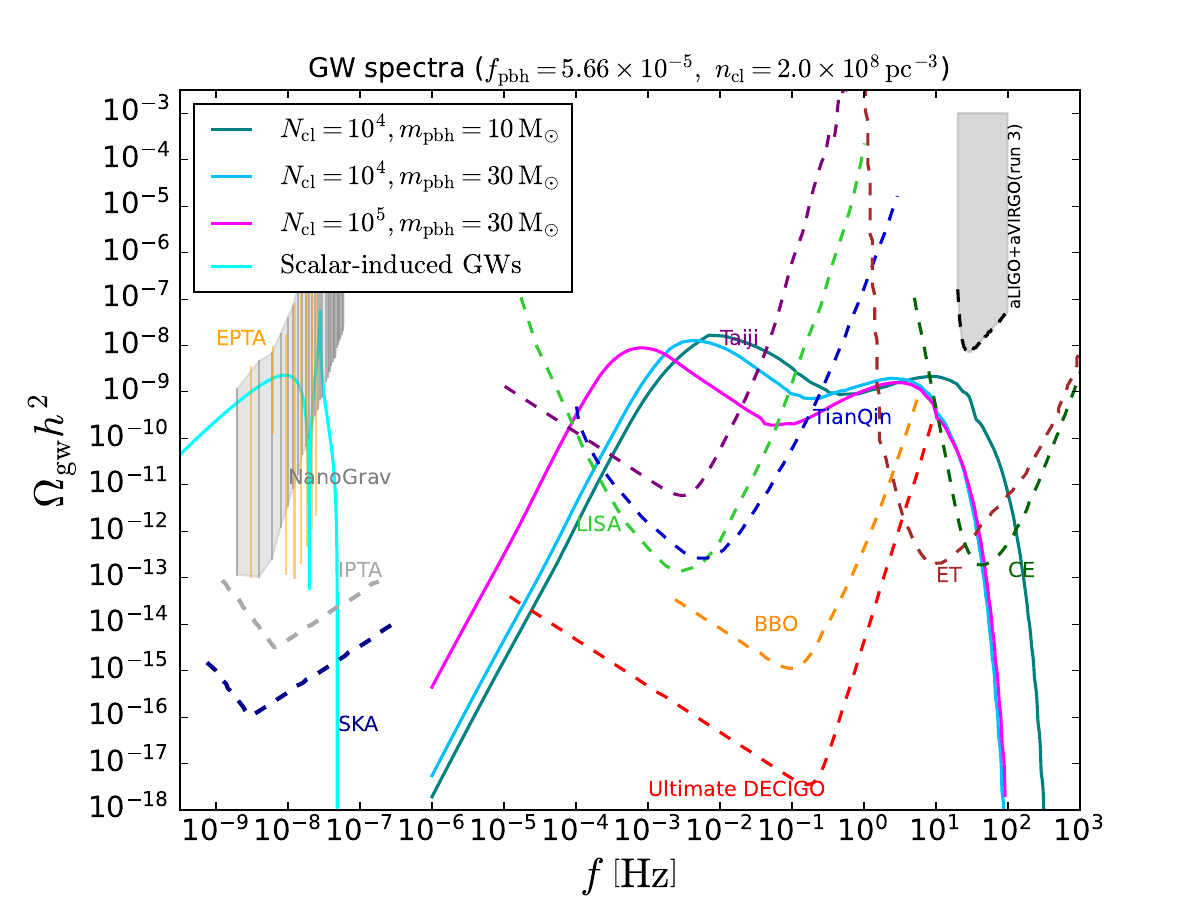}
	\caption{Stochastic GW background from PBH cluster merger processes and $\sim\mathcal{O}(10){\rm\,M}_\odot$ PBHs at formation. Sensitivities of future GW detectors\;\cite{Colpi:2024xhw,Ruan:2018tsw,TianQin:2020hid,Luo:2025ewp,Kudoh:2005as,Kawamura:2020pcg,Abac:2025saz,KAGRA:2021kbb} and PTA experiments\;\cite{Janssen:2014dka,Joshi:2025bvo,Verbiest:2016vem,NANOGrav:2023gor,NANOGrav:2023hvm,NANOGrav:2023hde,EPTA:2023sfo,EPTA:2023fyk,EPTA:2023xxk} are also shown.}
	\label{fig:stochastic-gw}
 \end{figure}

Fig.\,\ref{fig:stochastic-gw} shows the stochastic GW background associated with the PBH merging processes. A key feature of the GW spectrum is the presence of two peaks: one at low frequencies and the other at relatively high frequencies. The first peak arises due to the initially large abundance of small-mass PBHs in the first stage in Fig.\,\ref{fig:combined}. The third stage explains the emergence of the peak at low frequencies: The strength of the GW signal is proportional to a positive power of the chirp mass, while the maximum frequency is inversely proportional to the total mass. After the binary formation at relatively low redshifts, mergers of PBHs with their central SMBH contribute significantly at low frequencies, thereby enhancing the GW spectrum. The infrared behavior of the gravitational wave spectrum follows a $\sim f^3$ scaling, which is much steeper than the $f^{2/3}$ scaling of a single binary inspiral event. This is because orbital eccentricity transfers power from the fundamental harmonic to higher harmonics, significantly suppressing the GW background at frequencies below the source frequency while enhancing it at intermediate frequencies, see Supplemental Material for details. In addition to the merger processes, $\mathcal{O}(10){\rm\,M}_\odot$ PBHs at formation will accompany scalar-induced GWs\;\cite{An:2023jxf} with frequency $\sim10^{-9}\textup{--}10^{-8}{\rm\,Hz}$ in the band of Pulsar Timing Array (PTA) experiments\;\cite{Janssen:2014dka,Joshi:2025bvo,Verbiest:2016vem,NANOGrav:2023gor,NANOGrav:2023hvm,NANOGrav:2023hde,EPTA:2023sfo,EPTA:2023fyk,EPTA:2023xxk}.

\paragraph*{Accretion Growth and the Spin from Tidal Fields.---}
The mass $M_\bullet$ and spin $J_\bullet$ of the most massive black hole in the cluster are determined mainly by mergers and mass accretion. The accretion efficiency is sensitive to the spin of accreting black holes through the radiative efficiency $\epsilon_M=\epsilon_M(a_\bullet)$, which increases monotonically with the dimensionless spin, $a_\bullet\equiv J_\bullet/J_{\text{max}}$, and $J_{\rm max}\equiv GM_\bullet^2/c$ is the maximal spin of a Kerr black hole\;\cite{Bardeen:1972fi,Shapiro:2004ud}.  For example, $\epsilon_M(1)=0.42$, while $\epsilon_M(0.95)=0.19$, thus a small difference in spin results in a substantial difference in radiative efficiency. The enhancement of mass growth together with Eddington accretion is given by\;\cite{Shapiro:2004ud}
\begin{align}
	\frac{M_\bullet}{m_{\rm pbh}}=\mathcal{F}_{\rm merg}\,\mathrm{exp}\left[\frac{\epsilon_L\left(1-\epsilon_M\right)}{\epsilon_M\,\tau_{\rm acc}}t_{\rm acc}\right]\,,
\end{align}
where $\mathcal{F}_{\rm merg}\propto N_{\rm cl}$ is the merger enhancement factor, $\epsilon_L$ the accretion luminosity efficiency, $\tau_{\rm acc} \approx 0.4\,\text{Gyr}$ the characteristic accretion timescale, and $t_{\rm acc}$ the accretion time of baryon gas. 

For Eddington-limited accretion ($\epsilon_L=1$), the spin can be driven to its maximal equilibrium value ($a_\bullet\simeq1$ for a thin disk; $a_\bullet\simeq0.95$ for an MHD disk) through gas accretion in $t_{\rm acc}\sim0.1\,\tau_{\rm acc}\approx40{\rm\,Myr}$, and the mass growth is limited to \emph{at most} $10^4$ assuming a $100\%$ duty cycle in $t_{\rm acc}\sim2\,\tau_{\rm acc}\approx800{\rm\,Myr}$ for the MHD disk model ($\epsilon_M=0.19$). While the mass enhancement is only order unity if the duty cycle is $\sim10\%$\;\cite{Zhang:2026uum}. Thus, a sufficiently heavy seed is favored at the beginning of the accretion phase to explain those LRDs hosting \(10^{5\textup{--}8}{\rm\,M}_{\odot}\) SMBHs in the first billion years. Crucially, a slower mass growth rate can also reconcile the cumulative mass density accreted onto black holes with the SMBH mass density at \(z \simeq 4\textup{--}5\), based on the Sołtan argument\;\cite{Soltan:1982vf}, if their bolometric luminosities are estimated via spectral energy distribution fitting employing dust‑obscured AGN models\;\cite{Inayoshi:2024xwv}; while slowly-spinning SMBHs\;\cite{Pacucci:2024tws} are suggested to explain the X-ray weakness of LRDs. In fact, the spin evolution may depend on complex accretion models\;\cite{Ricarte:2023owr}. Chaotic accretion of short-lived episodes with random orientations tends to damp the black hole spin down to average values of $a_\bullet\simeq0.2$\;\cite{King:2008au}; cosmological hydro-simulations show that the gas angular momentum directions of subsequent accretion episodes become uncorrelated with the black hole spin once $M_\bullet\gtrsim2\times10^7{\rm\,M}_\odot$\;\cite{Sala:2023uxf}.
 
Regardless of accretion, our model suggests SMBHs with large spin, even if the PBHs had zero spin at their formation\;\cite{DeLuca:2019buf,Yoo:2024lhp}. 
The PBH cluster acquires significant spin through interactions with tidal fields induced by other clusters, due to its extended merger history and non-sphericity. As a result, the spin of the SMBH reflects the residual spin of the cluster after mergers. To estimate the magnitude, we consider the quadrupole interaction model\;\cite{Peebles:1969jm} and a homogeneous ellipsoid model for clusters, as spin deviations from homogeneity are only order unity. The dimensionless spin parameter of the cluster is estimated as (see Supplemental Material):
\begin{align}
&a_{\rm cl}\equiv cJ_{\text{cl}}/GM_{\text{cl}}^2\simeq5.6\,\mathcal{H}(e,p) \notag\\
\times&\left(\frac{f_{\rm pbh}}{10^{-5}}\right)^{\!-1/2}\!\!\left(\frac{\xi_{\alpha}}{10^6}\right)^{\!-1}\!\!\left(\frac{C}{20}\right)^{\!-1/6}\!\!\left(\frac{r_{\text{cl}}}{10\,\text{kpc}}\right)^{\!-1},
\end{align}
where $\mathcal{H}(e,p)$, with the ellipticity $e$ and prolateness $p$, characterizes the non-sphericity ($\mathcal{H}(0,0)=0$ if spherical) of the cluster according to peak theory\;\cite{Bardeen:1985tr,Sheth:1999su,DeLuca:2019buf,Escriva:2024lmm,Yoo:2024lhp}, $\mathcal{H}(\bar{e}\simeq0.09,\bar{p}\simeq0)\simeq0.53$ on average.

The light-gray region in Fig.\,\ref{fig:constraint} denotes the clusters with runaway timescale \(t_{\rm ra}\lesssim1{\rm\,Gyr}\), corresponding to \(z\gtrsim5\), to explain the SMBHs in LRDs.  In this region, the PBH cluster can acquire a large spin exceeding the Kerr limit. If successive mergers tend to align with the cluster spin, we find an asymptotic SMBH spin \(0.8\lesssim a_\bullet\lesssim0.95\)\;\cite{Buonanno:2007sv}, and the remaining PBHs then form extreme mass-ratio inspirals with the central SMBH. The darker band corresponds to \(t_{\rm ra}\gtrsim400{\rm\,Myr}\), while still allowing rapid SMBH formation by the relevant LRD redshifts.

\paragraph*{Discussion.---}
In addition to the requirements discussed in the preceding sections, the green region in Fig.\,\ref{fig:constraint} shows the experimental constraints on PBH abundance, $f_{\rm pbh}$\;\cite{Carr:2016drx,Sasaki:2018dmp,Carr:2020gox}, derived from gravitational lensing, Lyman-\(\alpha\) measurements, X-ray observations\;\cite{Inoue:2017csr}, and dynamical friction\;\cite{Carr:1997cn}.
As we focus on PBHs with an initial mass of \(\mathcal{O}(10){\rm\,M}_{\odot}\), their abundance is only weakly constrained, with \(f_{\rm pbh}\lesssim10^{-2}\). Especially, following successive mergers of PBHs in the cluster, their comoving number density $\bar{n}_{\rm pbh}\approx f_{\rm pbh} \left(30{\rm\,M}_\odot/m_{\rm pbh} \right){\rm\,kpc^{-3}}$ declines significantly over time, making them less subject to these observational constraints. While the comoving number density of all clusters $\bar{n}^c_{\rm cl}=\bar{n}_{\rm pbh}/N_{\rm cl}\approx10^4 f_{\rm pbh}(3\times10^6{\rm\,M}_\odot/N_{\rm cl}m_{\rm pbh}){\rm\,Mpc^{-3}}$\;\cite{DeLuca:2022bjs,DeLuca:2022uvz} remains roughly constant if they do not merge with each other; this may explain the constant comoving abundance of LRDs over a wide range of redshifts\;\cite{Zhang:2026uum}. The primary constraint on the PBH abundance then arises only from the SMBH mass density of $\rho_\bullet\sim10^5{\rm\,M}_\odot{\rm\,Mpc^{-3}}$ in the local Universe, currently accounting for $f_{\rm pbh}\sim10^{-5}$ dark matter density—consistent with our parameter choice. However, a more dedicated study is required, as some of the mass density may result from baryon accretion; while some of the PBH clusters could evaporate if they happen to be isolated (see Supplemental Material).

In Fig.\,\ref{fig:constraint}, we only present an average configuration of the long-mode field \(\Phi_l\) (given the parameters $|\eta|$ and $k_l/k_s$), which determines the typical PBH clusters and associated runaway timescales such that the SMBHs can emerge around or earlier than the observed LRD redshifts. In Ref.\,\cite{Zhang:2026uum}, we demonstrate that even for a distribution of $\Phi_l$, our model can still produce the SMBH mass function that agrees with those in LRDs. The most stringent constraints on \(\Phi_l\) arise from CMB spectral distortions\;\cite{Chluba:2012we,Chluba:2012gq,Kawasaki:2012kn,Kohri:2014lza,Nakama:2017xvq}, see also Refs.\,\cite{Sasaki:2018dmp,Carr:2020gox}. COBE/FIRAS observations\;\cite{Fixsen:1996nj} impose upper limits on the \(\mu\) and \(y\) distortion parameters, with \(|\mu| \lesssim 9 \times 10^{-5}\) and \(|y| \lesssim 1.5 \times 10^{-5}\). These constraints can be evaded through the long-short mode coupling. In our model, SMBHs do not form from the direct collapse of long-wavelength perturbations but instead emerge from successive mergers of small-mass PBHs. Crucially, the variance of long modes in our scenario is sufficiently small, \(\sigma_{l} \lesssim0.0064\), allowing the \(\mu\)-distortion constraint to be satisfied while maintaining a sufficient PBH abundance $f_{\rm pbh}\sim10^{-5}$, which will ultimately result in the SMBH population. The \(y\)-distortion constraint (purple in Fig.\,\ref{fig:constraint}) is weaker than that from \(\mu\)-distortions in the mass range $10^{5\textup{--}8}{\rm\,M}_{\odot}$ for LRDs, while it primarily excludes PBH clusters that would result in ultra-massive black holes, \(M_\bullet \gtrsim10^{11}{\rm\,M}_{\odot}\).

In the simulation, the runaway merger time of the PBH cluster is conservative, as we have neglected the spatial enhancement of merger kernel due to mass segregation as a result of dynamical friction. If this is considered\;\cite{Zhang:2026vjk}, the runaway merger can be even faster, as low-velocity massive PBHs will sink toward the center, and increase the merging cross section. This indeed prevents the relativistic instability of core collapse\;\cite{Feng:2021rst}, while it facilitates the merger and further relaxes the parameter space constraints in Fig.\,\ref{fig:constraint}. 

Alternative explanation of LRD spectral signatures may result from extreme stellar densities\;\cite{Baggen:2024vef}, and these purely dense star clusters are short-lived ($\lesssim1{\rm\,Myr}$), eventually resulting massive black holes, $M_\bullet\sim10^4{\rm\,M}_\odot$, through runaway stellar collisions\;\cite{Pacucci:2025ojp}. Whereas how those dense stellar structure originate remains unexplained. Our small-scale PBH clusters form during the radiation epoch and the runaway merger can ensue with SMBHs even in the dark ages, and thus precede the star formation. Moreover, the seed effect\;\cite{Carr:2018rid,Su:2023jno,Dayal:2025aiv} of PBHs at their formation can also bound other dark matter and baryons in the subsequent evolution, which may explain the dense gas in LRDs\;\cite{Inayoshi:2024aa,Casey:2025aa,Naidu:2025rpo,Rusakov:2025aa,Graaff:2025aa} and prevent evaporation of isolated PBH clusters. Despite the intricate dynamics of stellar evolution, GWs from the dark ages may help distinguish our model from stellar mergers.

\paragraph*{Conclusions.---}
Initial small-scale PBH clustering could potentially explain the rapid formation of SMBHs within the first billion years after the Big Bang, particularly the LRDs observed in recent JWST data. We have considered a simplified model in which small-scale clustering arises from long-wavelength mode modulation at the PBH formation scale. In this scenario, a central SMBH emerges through successive mergers of light PBHs within the cluster in a runaway manner. In particular, we compute the associated stochastic GW background of this runaway process, which encodes the population evolution of PBHs in the cluster. Due to tidal interactions from other clusters, the resulting SMBH is expected to be highly spinning, and extreme mass ratio inspiral systems with the central black hole will naturally form\;\cite{Feng:2024obn}. The GW signature associated with this mechanism may be detected by next-generation GW observatories such as the Einstein Telescope\;\cite{Hild:2008ng,Hild:2011np,Singh:2021zah}, and are distinguishable from alternative SMBH formation models, e.g., heavy PBH seeds\;\cite{DeLuca:2022bjs,Hooper:2023nnl,Qin:2025ymc,Guo:2026cuv,Allegrini:2026jqt} or gravothermal core collapse\;\cite{Feng:2020kxv,Roberts:2024wup,Jiang:2025jtr,Shen:2025evo,Feng:2025rzf,Roberts:2025poo}.

\bigskip
We thank Simeon Bird, Roberto Maiolino, Shi Pi, Javier Rubio, and Wenzer Qin for correspondence, as well as Junwu Huang and Huangyu Xiao for discussion.
This work is supported in part by the National Science Foundation of China (NSFC) under Grant Nos.\,12525506 and 12475107, the National Key R\&D Program of China under Grants Nos.\,2021YFC2203100 and 2017YFA0402204, the China Postdoctoral Science Foundation under Grant No.\,2024M761594, and the Shuimu Tsinghua Scholar Program.

\bibliographystyle{utphys}
\bibliography{LRDs}


\bigskip

\begin{center}
    \textbf{Appendix}
\end{center}

\begin{figure}[h]
\centering
\includegraphics[width=0.4\textwidth]{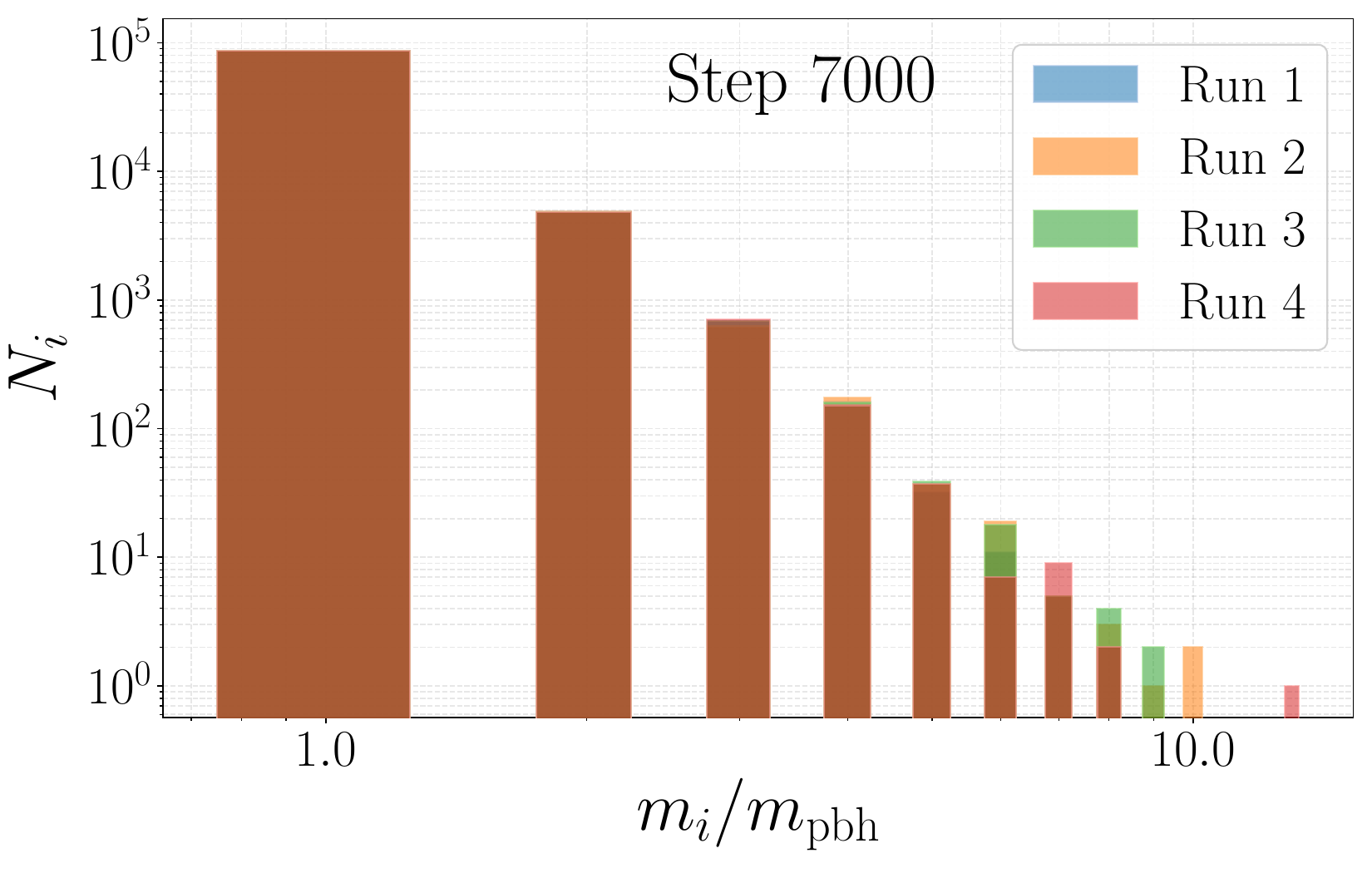}
\includegraphics[width=0.4\textwidth]{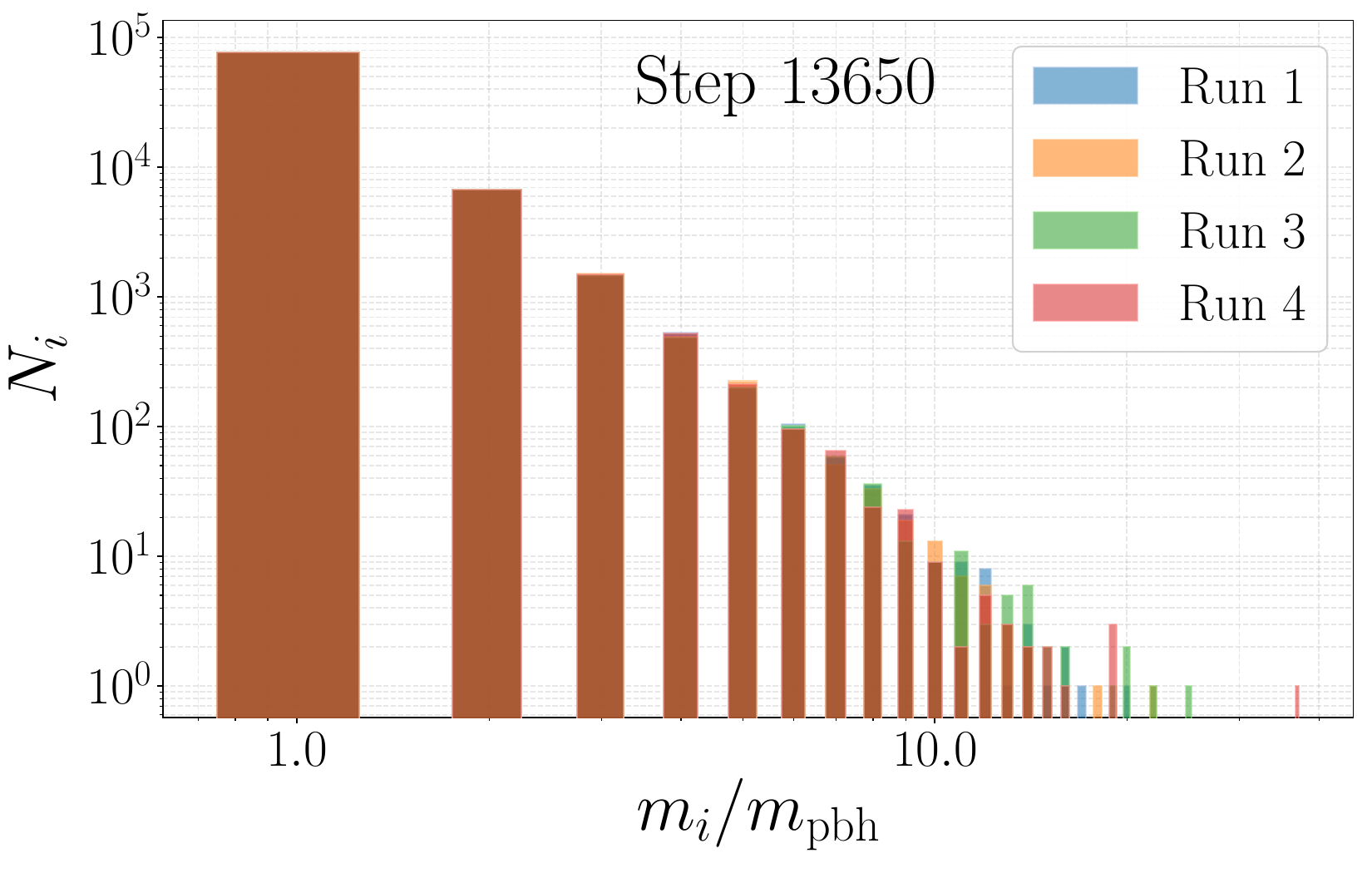}
\includegraphics[width=0.4\textwidth]{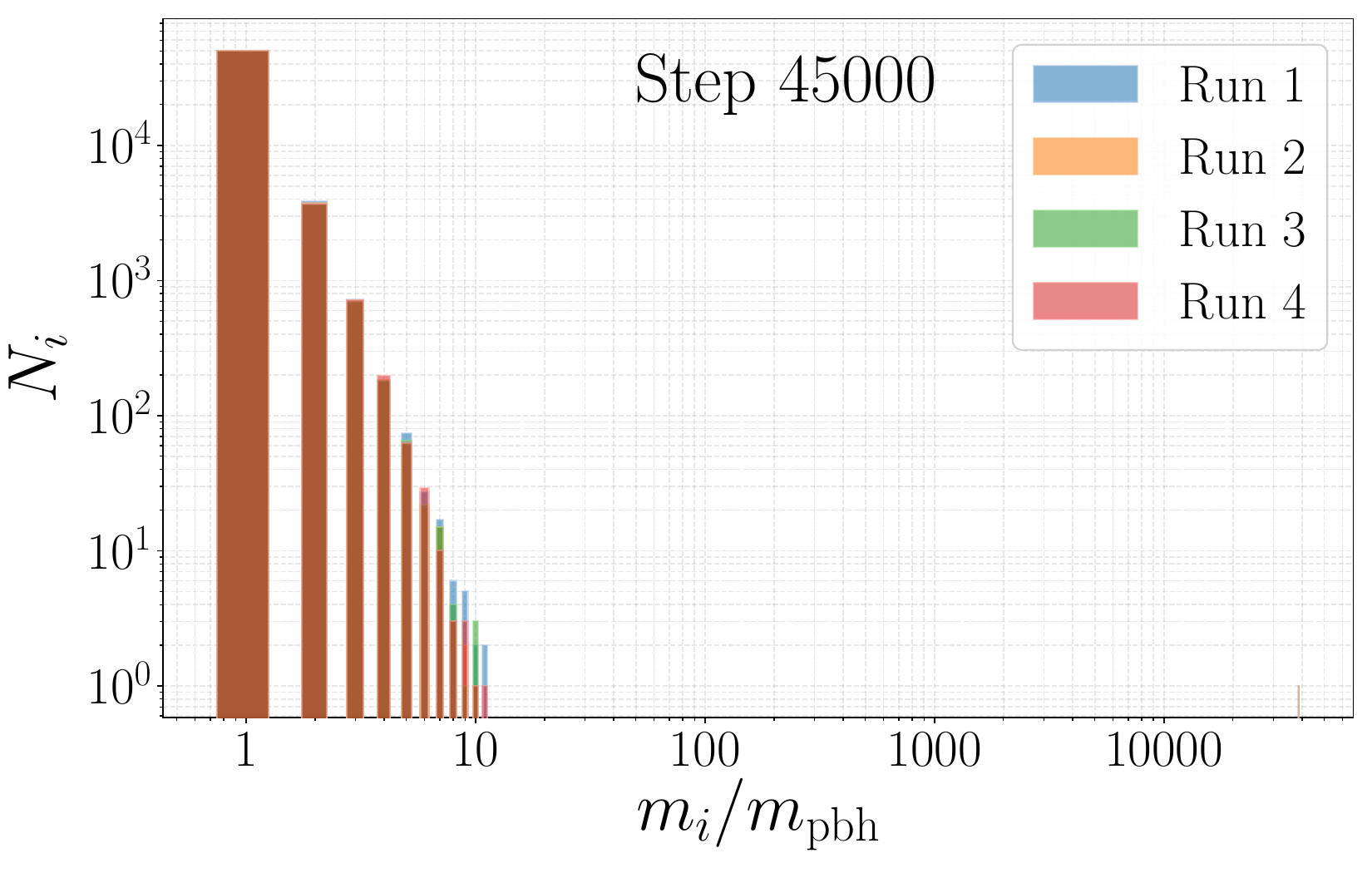}
\caption{Four different runs of the PBH mass population in a cluster with \(N_{\rm cl}=10^5\) for three evolution stages.}
\label{fig:combined_MC_result}
\end{figure}

\textit{Stability of the Monte Carlo simulation.}---
To show the stability of our Monte Carlo simulations, we perform four different runs of the PBH cluster evolution. Fig.\,\ref{fig:combined_MC_result} shows the cluster evolution for three stages corresponding to Fig.\,\ref{fig:combined}. Although there are small fluctuations when the number of PBHs is small at given masses, the Monte Carlo simulation is stable in the majority of PBHs and the mass of the central SMBH after runaway merger.

\textit{GWs from mergers in PBH clusters.}---
As orbital eccentricity modifies both the amplitude and the spectral shape of the GWs\;\cite{Enoki:2006kj,Clesse:2016ajp,Cholis:2016kqi}, we must trace the evolution of each PBH pair $(i,j)$ to compute the spectrum. The orbital dynamics of a binary system is characterized by the time evolution of its semimajor axis $\texttt{a}(t)$ and eccentricity $\texttt{e}(t)$\;\cite{Maggiore:2007ulw}. Through two-body capture, the initial semimajor axis and eccentricity are given, respectively, by
\begin{align}
\texttt{a}_0&=\frac{G(m_i+m_j)\mu_{ij}}{2|E_{{\rm b},ij}|};\notag\\
~\texttt{e}_0&=\sqrt{1+\frac{2E_{{\rm b},ij}v_{{\rm rel},ij}^2b_{ij}^2}{G^2(m_i+m_j)^2\mu_{ij}}}\,,
\end{align}
with
\begin{align}
E_{{\rm b},ij}=-\frac{1}{2}\mu_{ij}v_{{\rm rel},ij}^2\left[\left(\frac{b_{{\rm max},ij}}{b_{ij}}\right)^7-1\right]
\end{align}
the binding energy of the formed binary and $b_{{\rm max},ij}$ defined via the merging cross section $\sigma_{{\rm merg},ij}=\pi b_{{\rm max},ij}^2$ in Eq.\,\ref{eq:merg_sigma}, and $\mu_{ij}\equiv m_im_j/(m_i+m_j)$ the reduced mass.

By sampling the impact parameter $b_{ij}$ and relative velocity $v_{{\rm rel},ij}$, we obtain the initial $(\texttt{a}_0, \texttt{e}_0)$ distributions for each merging PBH pair. We assume that the impact parameter follows a uniform distribution within the range \( b_{ij} \in [0, b_{{\rm max},ij}] \); while the relative velocity follows a Maxwellian distribution given by the normalized probability density\;\cite{Mouri:2002mc,Zhang:2026vjk}:
\begin{align}
P(v_{{\rm rel},ij},\alpha_{ij})=\sqrt{\frac{2\alpha_{ij}^3}{\pi}}v_{{\rm rel},ij}^2\exp\left(-\frac{\alpha_{ij} v_{{\rm rel},ij}^2}{2}\right)
\end{align}
where $\alpha_{ij}=3\mu_{ij}/m_0v_0^2$ with $m_0=m_{\rm pbh}$.

According to Kepler's third law, we can associate the instantaneous orbital frequency with the emitted GW frequency $f_{ij}$ of the inspiral phase. The minimum GW frequency is twice of the orbital frequency at $\texttt{a}_0$:
\begin{align}
f_{{\rm min},ij}=2f^{(s)}_{{\rm min},ij}=\frac{1}{\pi}\sqrt{\frac{G(m_i+m_j)}{\texttt{a}_0^3}}\;,
\end{align}
and the maximum is determined at the innermost stable orbit (ISCO):
\begin{align}
f_{\mathrm{max},ij}=2f^{(s)}_{{\rm max},ij}\simeq \frac{c^3}{6^{3/2}G\pi \left(m_i+m_j\right)}\,,
\end{align}
where superscript ``${(s)}$'' denotes the orbital  quantities.

Following the standard procedure in Ref.\,\cite{Phinney:2001di} (see also Supplemental Material for details), we can compute the stochastic GW background energy fraction of present-day critical density due to the mergers in PBH clusters,
\begin{align}
&\Omega_{\rm gw}(f)
=\frac{\left(G\pi f\right)^{2/3}\bar{n}_{\rm cl}^c}{3\,\rho_cc^2}\int^{\infty}_0{\rm d}z\,(1+z)^{-1/3}\notag\\
&\times\sum_{ij}\mathcal{M}_{ij}^{5/3} N_{ij}(z)\,\Upsilon\left[\texttt{e},f,z,m_i,m_j\right]
\end{align}
where $\bar{n}_{\rm cl}^c=\bar{n}_{\rm pbh}/N_{\rm cl}$ is the average comoving number density of PBH clusters with $\bar{n}_{\rm pbh}\approx f_{\rm pbh} \left(30{\rm\,M}_\odot/m_{\rm pbh} \right){\rm\,kpc^{-3}}$, $\rho_c=\rho_{\rm crit,0}$ is the present-day critical density, $\mathcal{M}_{ij}=\left(m_im_j\right)^{3/5}/\left(m_i+m_j\right)^{1/5}$ denotes the chirp mass, $N_{ij}(z)$ the number of merger events for a given pair $(i,j)$ occurring at redshift $z$, and
\begin{align}
&\Upsilon(\texttt{e},f,z,m_i,m_j)=\sum_{n=1}^{\infty}\left(\frac{n}{2}\right)^{-2/3}\frac{\mathcal{G}(n,\texttt{e})}{\mathcal{F}(\texttt{e})}\notag\\
&\times\theta\left[\ln\left(\frac{nf_{{\rm max},ij}}{2(1+z)}\right)-\ln f\right]\,\theta\left[\ln f-\ln\left(\frac{nf_{{\rm min},ij}}{2(1+z)}\right)\right]\,,
\end{align}
with $\theta(x)$ the step function, and the functional forms of $\mathcal{G}(n,\texttt{e})$, $\mathcal{F}(\texttt{e})$ and
\begin{align*}
\texttt{e}=\texttt{e}\left(\frac{(1+z)f}{n},\texttt{e}_0,\texttt{a}_0\right)
\end{align*}
detailed in  Supplemental Material.
Here the observed frequency $f$ is related by $f=f_r/(1+z)=nf^{(s)}_n/(1+z)$, where $f^{(s)}_n$ denotes the orbital frequency corresponding to the $n$-th harmonic at rest-frame frequency $f_r$.

\textit{The associated scalar-induced GWs.}---
In addition to the GWs produced by binary mergers in a cluster, the associated scalar-induced GWs can also provide a distinctive signature of our model. PBHs formed through large primordial curvature perturbations $\zeta$, at horizon reentry of the scale wavenumber $k$, will induce strong second-order GWs. For PBHs of $\sim\mathcal{O}(10)\,\mathrm{M}_\odot$, whose comoving momentum $k_\star\sim\mathcal{O}(1)\,{\rm pc^{-1}}$, the GW frequencies peak around $f_\star=k_\star c\sim 10^{-8}{\rm\,Hz}$ within the band of PTA experiments\;\cite{Janssen:2014dka,Joshi:2025bvo,Verbiest:2016vem,NANOGrav:2023gor,NANOGrav:2023hvm,NANOGrav:2023hde,EPTA:2023sfo,EPTA:2023fyk,EPTA:2023xxk}. Therefore, we consider a monochromatic power spectrum that can be approximated by a delta function $\delta(x)$:
\begin{align}
\mathcal{P}_\zeta(k)\simeq \sigma_\zeta^2\,k_\star\,\delta(k-k_\star)\;.
\end{align}
The scalar-induced GWs can be computed as\;\cite{Kohri:2018awv,An:2023jxf}:
\begin{align}
&\Omega_{\rm gw}(k)=\sigma_\zeta^4\,\Omega_{\rm R}\,\frac{k_\star^2}{k^2}\frac{\nu_\star^3}{6}\left[1-\left(\frac{1}{4\nu^2_\star}\right)^2\right]^2\frac{\overline{I^2}(1,\nu_\star,\nu_\star)}{\nu_\star^3}
\end{align}
with $\Omega_{\rm R}$ the present-day radiation energy fraction and
\begin{align}
&\overline{I^2}(1,\nu_\star,\nu_\star)=\frac{1}{2}\left[\frac{3(2\nu_\star^2-3)}{4\nu_\star^6}\right]^2\bigg\{\pi^2(2\nu_\star^2-3)^2\,\theta(2k_\star/\sqrt{3}-k)\notag\\
&+\left[-4\nu_\star^2+(2\nu_\star^2-3)\ln\bigg|\frac{3-4\nu_\star^2}{3}\bigg|\right]^2\bigg\}\,\theta(2k_\star-k)\,,
\end{align}
where $\nu_\star\equiv k_\star/k$, $\mu_\star\equiv1/(2\nu_\star)$. In Fig.\,\ref{fig:stochastic-gw}, we set the variance of the curvature perturbation $\sigma_\zeta\sim0.1$.

\appendix
\clearpage
\onecolumngrid
\setcounter{page}{1}

\begin{center}
\textbf{\large --- Supplemental Material ---\\ $~$ \\
Little Red Dots from Small-Scale Primordial Black Hole Clustering}\\
\medskip
\text{Borui Zhang, Wei-Xiang Feng, and Haipeng An}
\end{center}
\setcounter{equation}{0}
\makeatletter
\renewcommand{\thesection}{S.\arabic{section}}
\renewcommand{\theequation}{S.\arabic{section}--\arabic{equation}}

\section{The two-point correlation function of PBHs}
A simple parametrization of the long-mode modulation is given by\;\cite{Atal:2020igj}:
\begin{align}
\nu(\mathbf{x})=\nu_{g}\left(1+\eta\,\Phi_{l}(\mathbf{x})\right)\,,
\end{align}
where \(\eta\) denotes the coupling strength between long and short modes, and \(\Phi_l(\mathbf{x})\) is the long-wavelength density perturbation field. The unmodulated reduced threshold is defined as \(\nu_g \equiv \delta_c / \bar{\sigma}_{s}\) with \(\delta_c \simeq 0.414\) the critical density contrast for PBH formation and \(\bar{\sigma}_{s}^2 = \langle\delta_s^2\rangle\) the spatially averaged variance of the short-mode density perturbations. Both $\Phi_l$ and $\delta_s$ are assumed to follow Gaussian statistics. 
The probability of forming a PBH and the joint probability of finding two PBHs separated by a distance $r$, after integrating over the configuration of the long mode are, respectively,
\begin{align}
P_1=\frac{1}{2}\mathrm{erfc}\left(\frac{\nu(\mathbf{x})}{\sqrt{2}}\right)&
\quad{\rm and}\quad
P_2=\frac{1}{4} \left[ \text{erfc} \left( \frac{\nu_1}{\sqrt{2}} \right) + \text{erfc} \left( \frac{\nu_2}{\sqrt{2}} \right) + \text{sgn}(\nu_1) \text{sgn}(\nu_2) - 1 \right] \nonumber\\
    &-T \left( \nu_1, \frac{\nu_2 - \omega_s(r)\nu_1}{\nu_1\sqrt{1- \omega_s^2(r)}} \right)
    -T \left( \nu_2, \frac{\nu_1 - \omega_s(r)\nu_2}{\nu_2\sqrt{1- \omega_s^2(r)}} \right)\,,
\end{align}
where $\nu_i = \nu(\mathbf{x}_i)$ is the local reduced threshold, $\omega_s(r) = \langle \delta(\mathbf{x}_1) \delta(\mathbf{x}_2) \rangle/\langle \delta(0)^2 \rangle$ the normalized two-point correlation function of the short-wavelength density field, $\text{erfc}(x)\equiv\frac{2}{\sqrt{\pi}}\int_x^\infty e^{-t^2}{\rm d}t$, and $T(z, u)$ the Owen $T$-function, given by
\begin{equation}
T(z, u) = \frac{1}{2\pi} \int_0^u \frac{e^{-\frac{(1+t^2)z^2}{2}}}{1+t^2}{\rm\,d}t\;.
\end{equation}

For this simple modulation model, the $N$-point correlation function of PBHs can be computed analytically\;\cite{Atal:2020igj}. In particular, the two-point correlation function $\xi_{\rm pbh}(r)$ of PBHs is given by
\begin{align}
    1+\xi_{\rm pbh}(r)=\frac{P_2(r)}{P_1^2}
    \stackrel{\nu\gg1}{\longrightarrow}\frac{\left(1+\bar{\omega}(r)\right)^{3/2}}{\left(1-\bar{\omega}(r)\right)^{1/2}}\mathrm{exp}\left(\nu_g^2\frac{\bar{\omega}(r)}{1+\bar{\omega}(r)}\right)\,,
\end{align}
where $\bar{\omega}(r)\equiv \frac{\omega_s(r)+\alpha^2\omega_l}{1+\alpha^2}$ with $\alpha(\eta,\kappa)\equiv \delta_c\eta\kappa$ and $\kappa\equiv\sigma_{l}/\bar{\sigma}_{s}$ denoting the ratio of long- to short-mode variances. For simplicity, we adopt a model for the primordial power spectrum composed of two delta functions:
\begin{equation}
\begin{cases}
\mathcal{P}_{s,\delta}(k)=\sigma_{s}^2k_s\delta(k-k_s)\,,\\
\mathcal{P}_{l,\Phi}(k)=\sigma_{l}^2k_l\delta(k-k_l)\,,
\end{cases}
\end{equation}
where $k_{s,l}$ represent the wavenumbers of the short and long modes, and $\sigma_{s,l}$ are the corresponding variances.

\begin{figure}[H]
    \centering
    \includegraphics[width=0.49\textwidth]{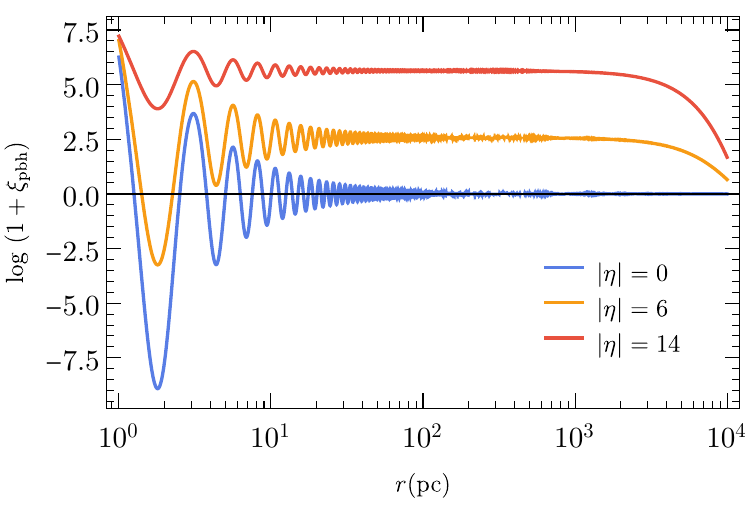}
    \caption{The PBH two-point correlation function for $\abs{\eta}=0,\,6$ and $14$ given $\nu_g=8.5$, $\kappa=0.1313$.}
    \label{fig:2pt}
\end{figure}

Fig.\,\ref{fig:2pt} displays the PBH two-point correlation function for different coupling strengths $\abs{\eta}=0,\,6$ and $14$, given \( \nu_g = 8.5 \), \( \kappa = 0.1313 \), and \( k_s \simeq 2.5{\rm\,pc}^{-1} \) corresponding to \( m_{\text{pbh}} = 30{\rm\,M}_{\odot} \). We set the ratio of long to short wavelengths as \( k_l / k_s=10^{-4}\). Two distinct features are observed in the correlation function. The first drop occurs at the characteristic scale of an individual PBH, \( r_{\text{pbh}} \sim 1/k_s \), while the second drop appears at the typical clustering scale set by the long mode, \( r_{\text{cl}} \sim 1/k_l \). When \( |\eta| \neq 0 \), a plateau emerges in the intermediate regime \( r_{\text{pbh}} < r < r_{\text{cl}} \), where $\xi_{\rm pbh}(r)$ remains nearly constant at $\xi_\alpha$. This plateau reflects the enhancement of PBH clustering induced by the long-mode modulation. 

\section{The total number of PBHs and the number density in a cluster}
The total number of PBHs in a cluster can be estimated using the two-point correlation function\;\cite{DeLuca:2022bjs,DeLuca:2022uvz}:
\begin{equation}
N_{\text{cl}} \approx \bar{n}_{\rm pbh} \int{\rm d}^3x\,\xi_{\rm pbh}(r) \approx \frac{4\pi}{3} \bar{n}_{\rm pbh} \xi_\alpha r_{\text{cl}}^3
\end{equation}
where $\bar{n}_{\rm pbh}$ is the average comoving number density of PBHs:
\begin{equation}
\bar{n}_{\rm pbh} 
=\frac{\Omega_{\rm dm}}{\Omega_{\rm M}}f_{\rm pbh}\frac{\rho_{\rm eq}}{m_{\rm pbh}}a_{\rm eq}^3
\approx f_{\rm pbh} \left( \frac{30{\rm\,M}_\odot}{m_{\rm pbh}} \right) \text{kpc}^{-3}\;.
\end{equation}
Here, we have used the ratio of dark matter to total matter abundance $\Omega_{\rm dm}/\Omega_{\rm M}\approx 0.85$, the PBH fraction of dark matter $f_{\rm pbh}\equiv\Omega_{\rm pbh}/\Omega_{\rm dm}$, and the matter energy density at matter-radiation equality $\rho_{\rm eq}\equiv\Omega_{\rm M}\,\rho_{\rm crit,0}/a_{\rm eq}^3$, with the scale factor $a_{\rm eq}\simeq1/3401$. The present critical density is taken to be $\rho_{\rm crit,0}\simeq1.27\times10^{-7}{\rm\,M}_\odot{\rm\,pc^{-3}}$ and $\Omega_{\rm M}\simeq0.3$.

We consider the PBH cluster to evolve as normal matter, diluting with the cosmic expansion until it decouples from the Hubble flow. This decoupling occurs when the energy density of the cluster, $\rho_{\rm cl}=m_{\rm pbh}n_{\rm cl}$, becomes comparable to the radiation energy density, $\rho_{\rm rad}$. After this point, the number density $n_{\rm cl}$ becomes constant in comoving coordinates. The cluster's energy density at decoupling, with scale factor $a_{\rm dec}$, is given by:
\begin{equation}
\rho_{\text{cl}}(a_{\text{dec}}) = \xi_\alpha \bar{\rho}_{\rm pbh}(a_{\text{dec}}) = \xi_\alpha \bar{\rho}_{\text{pbh}}(a_{\text{eq}}) \left( \frac{a_{\text{eq}}}{a_{\text{dec}}} \right)^3 = \xi_\alpha \left(0.85f_{\rm pbh} \rho_{\text{eq}}\right) \left( \frac{a_{\text{eq}}}{a_{\text{dec}}} \right)^3\,,
\end{equation}
where $\bar{\rho}_{\rm pbh}=m_{\rm pbh}\bar{n}_{\rm pbh}$. On the other hand, the radiation energy density redshifts as:
\begin{equation}
\rho_{\text{rad}}(a_{\text{dec}}) = \rho_{\text{eq}} \left( \frac{a_{\text{eq}}}{a_{\text{dec}}} \right)^4.
\end{equation}
Equating $\rho_{\text{cl}}(a_{\text{dec}})=\rho_{\text{rad}}(a_{\text{dec}})$, we find the scale factor at decoupling:
\begin{equation}
 a_{\text{dec}} = \frac{a_{\text{eq}}}{0.85f_{\rm pbh} \xi_\alpha}\,,
\end{equation}
which is valid for $0.85f_{\rm pbh} \xi_\alpha>1$. After decoupling and virialization, the cluster undergoes further contraction. We account for this by introducing the compactness parameter $C\gg1$, which enhances the final cluster number density:
\begin{equation}
n_{\text{cl}} = C(0.85f_{\rm pbh}\xi_\alpha)^4 \frac{\rho_{\text{eq}}}{m_{\rm pbh}}
\simeq2\times10^8{\rm\,pc^{-3}}\left(\frac{f_{\rm pbh}\xi_\alpha}{25}\right)^4\left(\frac{C}{20}\right)\left(\frac{30{\rm\,M}_{\odot}}{m_{\text{pbh}}}\right)\;.
\end{equation}

\section{The spin of clusters from tidal interactions}
We adopt a model in which PBH clusters are approximated as homogeneous ellipsoids with semi-axes $r_1\geq r_2\geq r_3$ during their hierarchical merging phase leading to SMBH formation.
The tidal torque yields
\begin{align}
&\tau_i=\epsilon_{ijk}Q_{jl}T_{lk}\,,
\end{align}
where $Q_{ij}$ is the quadrupole moment of the cluster, $T_{ij}=GM_{\rm cl}\left(\delta_{ij}-3\hat{x}_i\hat{x}_j\right)/r^3$ the tidal tensor of a source at $(r, \theta, \phi)$ with the unit vector $\hat{x}_i=(\sin\theta \cos\phi, \sin\theta \sin\phi, \cos\theta)$. The ensemble-averaged squared tidal torque is computed by integrating over all external cluster configurations
\begin{align}
\langle\tau^2\rangle=\int^{\infty}_{r_{\rm vir}}P(r)\left(\sum_{i=1}^3 \tau_i^2\right){\rm\,d}^3x\,,
\end{align}
where $P(r)$ is the two-point spatial distribution function of PBH clusters, describing their correlation at separation $r$, and is given as follows.

We assume that all PBH clusters have the same mass and are homogeneously distributed within a maximum comoving distance $R_{\rm max}^c$, which defines the local clustering environment. The comoving mean inter-cluster distance is given by
\begin{align}
\bar{R}_{\rm cl}^c
\simeq \left(\frac{4\pi\bar{n}_{\rm pbh}}{3N_{\rm cl}}\right)^{-1/3}=\xi_{\alpha}^{1/3}r_{\rm cl}\,,
\end{align}
where $\xi_\alpha\gg1$ ensures a high clustering factor, and hence $\bar{R}_{\rm cl}^c\gg r_{\rm cl}$. We set $R_{\rm max}^c\simeq(4/3)\bar{R}_{\rm cl}^c$ using the definition of the average distance $\bar{R}_{\rm cl}^c=\int_{|\mathbf{x}|<R_{\rm max}^c}|\mathbf{x}-\mathbf{x}_0|{\rm\,d^3}\mathbf{x}/\int_{|\mathbf{x}|<R_{\rm max}^c}{\rm d^3}\mathbf{x}$ for a reference cluster fixed at the origin $\mathbf{x}_0=0$.
Evaluating the physical separation at the decoupling epoch $a_{\rm dec}\simeq a_{\rm eq}/(0.85f_{\rm pbh}\xi_\alpha)$, we find:
\begin{align}
R_{\rm max}^p(a_{\rm dec})\simeq a_{\rm dec}\bar{R}_{\rm cl}^c=a_{\rm eq}/(0.85f_{\rm pbh}\xi_\alpha)\cdot \frac{4}{3}\xi_{\alpha}^{1/3}r_{\rm cl}=\frac{4}{3}\left(C\xi_\alpha\right)^{1/3}r_{\rm vir}\,,
\end{align}
where we have used the virial radius expression:
\begin{equation}
r_{\rm vir}=\left(\frac{3M_{\rm cl}}{4\pi \rho_{\rm cl}}\right)^{1/3}
=\left(\frac{3N_{\rm cl}}{4\pi n_{\rm cl}}\right)^{1/3}
=(0.85f_{\rm pbh}\xi_\alpha)^{-1}C^{-1/3}a_{\rm eq}\,r_{\rm cl}\;.
\end{equation}
Thus, the probability distribution for finding another cluster at position $\mathbf{x}_0+\mathbf{x}$ , given a cluster at $\mathbf{x}_0=0$, is approximately:
\begin{align}
P(r)\simeq
\left\{ 
\begin{matrix}
\frac{3}{4\pi}(R_{\rm max}^p)^{-3}\,,&r_{\rm vir}<r\lesssim R_{\rm max}^p\\	
0\,,&{\rm otherwise}
\end{matrix}
\right.\;.
\end{align}
Consequently, the ensemble-averaged tidal torque acting on a cluster is given by:
\begin{align}
\langle\tau^2\rangle^{1/2}=\frac{27}{40\sqrt{5}}\left(\frac{GM_{\rm cl}^2}{r_{\text{vir}}}\right)\left(C\xi_\alpha\right)^{-1/2}\left(\frac{a_{\rm dec}}{a}\right)^{3/2}\left[\left(1-\frac{r_2^2}{r_1^2}\right)^2+\left(1-\frac{r_3^2}{r_1^2}\right)^2+\left(\frac{r_2^2}{r_1^2}-\frac{r_3^2}{r_1^2}\right)^2\right]^{1/2}\,,
\end{align}
where we have used $R_{\rm max}^p(a_{\rm dec})\simeq(4/3)\left(C\xi_\alpha\right)^{1/3}r_{\text{vir}}$ and identified $r_{\text{vir}}=r_1$ as the major axis of the ellipsoid. Since $\langle\tau^2\rangle^{1/2}\propto a^{-3/2}$, the tidal torque weakens with cosmic expansion, indicating that most of the spin is accumulated shortly after the clusters decouple from the Hubble flow.
 
The final spin of a cluster can be estimated by integrating the ensemble-averaged tidal torque over the time interval during which the cluster remains dynamically responsive to the tidal field of neighboring clusters. 
The upper limit of this interval is approximated by the typical time before the cluster fully merges into SMBH:
\begin{align}
J_{\text{cl}}=\int^{t_{\text{merg}}}_{t_{\text{dec}}}\langle\tau^2\rangle^{1/2}{\rm\,d}t
\end{align}
where $t_{\rm dec}$ is the decoupling time and $t_{\rm merg}\lesssim t_{\rm ra}$ denotes the typical merging time. After performing the time-dependent integral
\begin{align}
\int^{t_{\rm merg}}_{t_{\rm dec}}\frac{a_{\rm dec}^{3/2}}{a^{3/2}}{\rm\,d}t
=\frac{2}{\sqrt{8\pi G\rho_{\rm eq}/3}}\left(0.85f_{\rm pbh}\xi_\alpha\right)^{-3/2}\ln\left(\frac{\sqrt{1+a_{\rm merg}/a_{\rm eq}}+\sqrt{a_{\rm merg}/a_{\rm eq}}}{\sqrt{1+1/(0.85f_{\rm pbh}\xi_\alpha)}+\sqrt{1/(0.85f_{\rm pbh}\xi_\alpha)}}\right)
\end{align}
with $a_{\rm eq}\simeq1/3401$, and $a_{\rm merg}\sim1/10$ around redshift $z\sim9$, we estimate the dimensionless spin parameter $a_s\equiv cJ_{\text{cl}}/GM_{\text{cl}}^2$ of the cluster as:
\begin{align}
a_s&\simeq5.6\left(\frac{f_{\rm pbh}}{10^{-5}}\right)^{-1/2}\left(\frac{\xi_{\alpha}}{10^6}\right)^{-1}\left(\frac{C}{20}\right)^{-1/6}\left(\frac{r_{\text{cl}}}{10\,\text{kpc}}\right)^{-1}
\left[\left(1-\frac{r_2^2}{r_1^2}\right)^2+\left(1-\frac{r_3^2}{r_1^2}\right)^2+\left(\frac{r_2^2}{r_1^2}-\frac{r_3^2}{r_1^2}\right)^2\right]^{1/2}\;.
\end{align}
While clusters are also subject to tidal torques induced by background density fluctuations, we neglect these contributions as they are subdominant compared to the torques from neighboring clusters for the parameter range considered.

According to the peak theory, the non-sphericity of the long-wavelength curvature perturbation can be characterized by its ellipticity $e$ and prolateness $p$, which follow a statistical distribution determined by the power spectrum of the initial perturbations and the peak value\;\cite{Bardeen:1985tr,Sheth:1999su,DeLuca:2019buf,Escriva:2024lmm,Yoo:2024lhp}.  
We assume a monochromatic mass distribution and calculate the corresponding peak value using the two-point correlation function. The axis ratios are then given by
\begin{align}
\frac{r_2}{r_1}=\sqrt{\frac{1-3\bar{e}+\bar{p}}{1+3\bar{e}+\bar{p}}}\,,\quad{\rm and}\quad \frac{r_3}{r_1}=\sqrt{\frac{1-3\bar{e}+\bar{p}}{1-2\bar{p}}}
\end{align}
with the average values $\bar{e}\simeq0.09$ and $\bar{p}\simeq 0$ as our inputs.

\section{Redshift evolution of the most massive black hole with various cluster densities}
 \begin{figure}[H]
	\centering
	\includegraphics[width=0.49\textwidth]{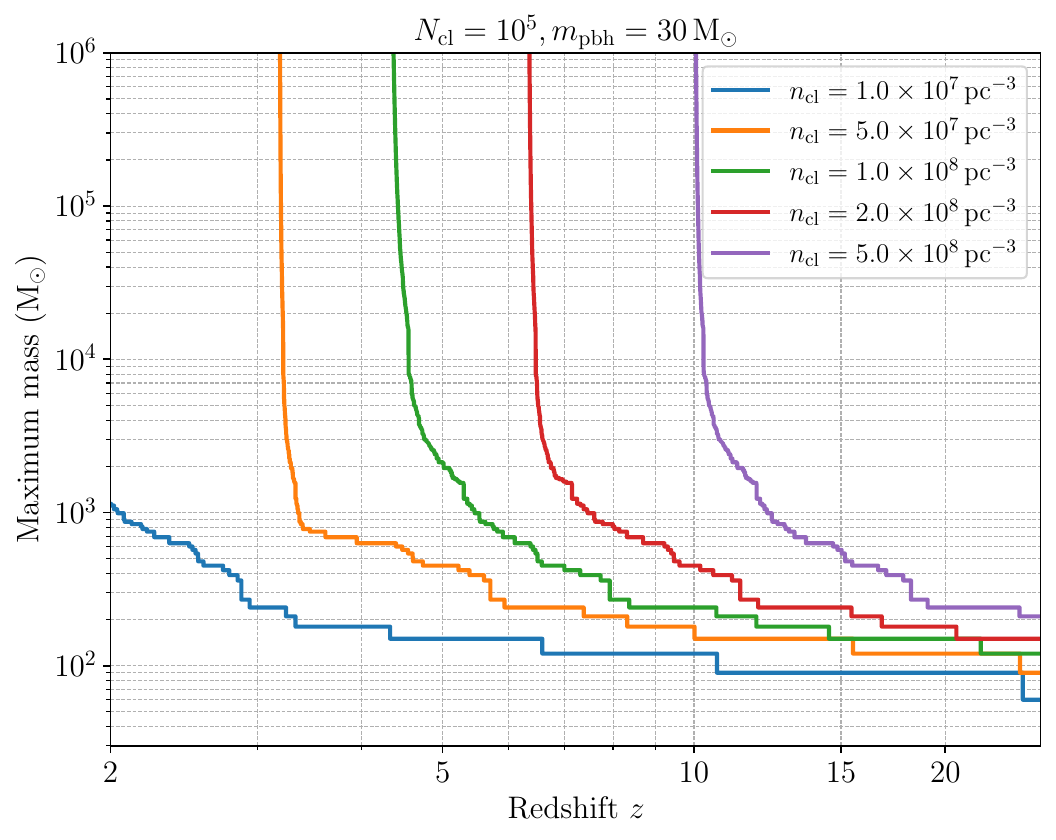}
	\includegraphics[width=0.49\textwidth]{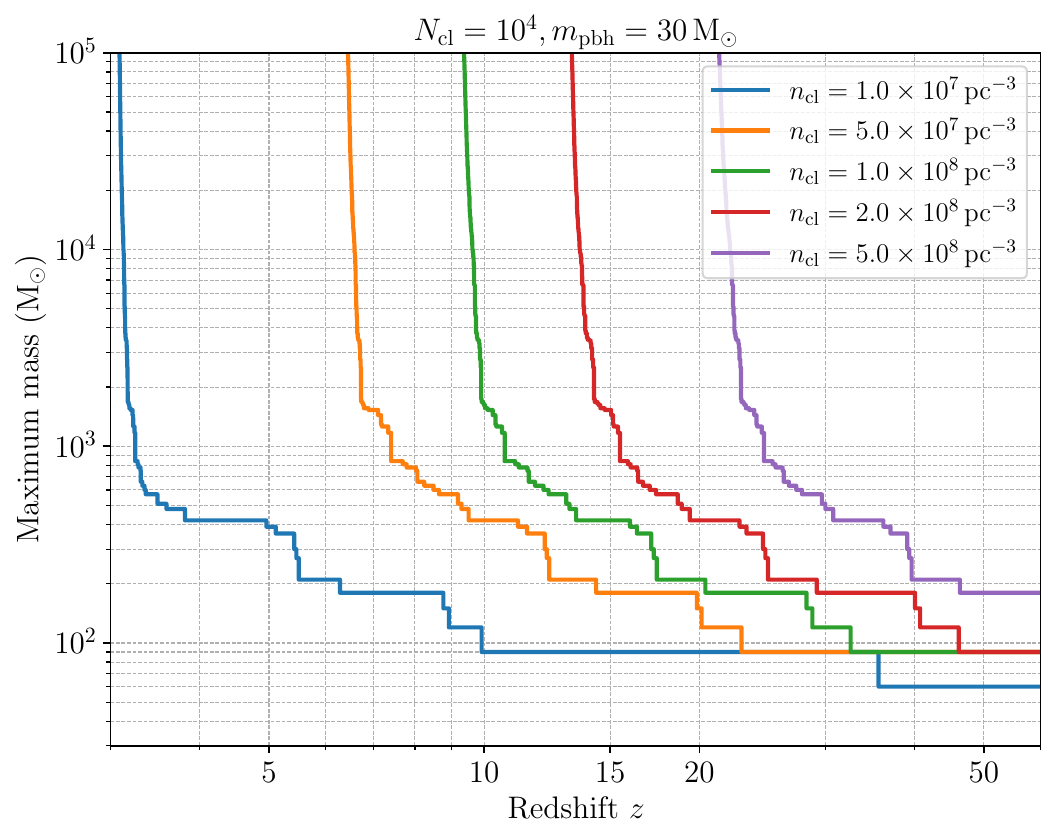}
	\caption{Redshift evolution of the most massive black hole in a PBH cluster of $N_{\rm cl}=10^5\,(10^4)$ PBHs of $m_{\rm pbh}=30{\rm\,M}_\odot$ with various initial cluster densities $n_{\rm cl}=(1,\,5)\times10^7$ and $(1,\,2,\,5)\times10^8{\rm\,pc^{-3}}$.}
	\label{fig:max_bh}
 \end{figure}

\section{Various timescales associated with a PBH cluster}
Once two PBHs form a binary, they will eventually coalesce due to the emission of GWs. However, this process can be interrupted by encounters with a third PBH within the same cluster. To safely neglect such three-body interactions in our Monte Carlo simulation, we require that the inspiral timescale be shorter than the typical disruption timescale due to a third-body interaction. The coalescence timescale due to GW emission,
\begin{equation}
t_{\text{gw}} \approx \frac{3c^5r_{\rm p}^4(1+\texttt{e})^{7/2}/(1-\texttt{e})^{1/2}}{85\,G^3(m_i+m_j)m_im_j}\,,
\end{equation}
must be shorter than the disruption timescale by a third PBH,
\begin{equation}
t_{\text{disrup}} \approx \frac{1}{n_{\text{cl}} \sigma_{\rm p} v_{\rm vir}}\,,
\end{equation}
where $\texttt{e}\lesssim1$ is the initial orbital eccentricity and $\sigma_{\rm p}\equiv\pi r_{\rm p}^2(1-\texttt{e})^{-2}$ the effective cross section of the newly formed binary with periapsis\;\cite{Feng:2024obn}
\begin{equation}
r_{\rm p} \approx \left( \frac{85\pi}{6\sqrt{2}} \frac{G^{7/2}(m_i + m_j)^{3/2} m_i m_j}{c^5v_{\rm vir}^2} \right)^{2/7}\;.
\end{equation}
It can be easily justified that given $\texttt{e}\simeq0.99$, $m_i\simeq m_j=m_{\rm pbh}=30{\rm\,M}_\odot$, $v_{\rm vir}=\sqrt{GN_{\rm cl}m_{\rm pbh}/r_{\rm vir}}$ and $n_{\rm cl}=3 N_{\rm cl}/4\pi r_{\rm vir}^3$ with $N_{\rm cl}=10^5$, we obtain $t_{\rm gw}/t_{\rm disrup}\gtrsim1$
only if $n_{\rm cl}\gtrsim10^{29}{\rm\,pc^{-3}}$; while in another extreme situation when SMBH emerges with $m_i=m_{\rm pbh}=30{\rm\,M}_\odot$ and $m_j\simeq10^6{\rm\,M}_\odot$, the result shifts to $n_{\rm cl}\gtrsim3\times10^{12}{\rm\,pc^{-3}}$. Both cases require much much larger densities than our typical cluster of $n_{\rm cl}\sim10^{7\textup{--}9}{\rm\,pc^{-3}}$ for the three-body disruption to be relevant.

Without mergers and losing energy through emitting GWs, an isolated PBH cluster will gradually shrink and evaporate when some PBHs undergo multiple gravitational scatterings during multiple (two-body) relaxation timescales to exceed the escape velocity, and this defines the evaporation timescale of a cluster:
\begin{equation}
t_{\text{evap}} \approx \Gamma\, t_{\rm relx}
\approx 12.7{\rm\,Myr} \left(\frac{\Gamma}{135}\right)\left(\frac{10}{\ln N_{\text{cl}}}\right) \left( \frac{N_{\text{cl}}}{10^5} \right)^{1/2} \left( \frac{m_{\rm pbh}}{30{\rm\,M}_\odot} \right)^{-1/2} \left( \frac{r_{\rm vir}}{0.05\rm\,pc} \right)^{3/2},
\end{equation}
where the relaxation time
\begin{equation}
t_{\text{relx}} \approx \frac{0.1\,N_{\text{cl}}}{\ln N_{\text{cl}}} \frac{r_{\rm vir}}{v_{\rm vir}}\,,
\end{equation}
and $1/\Gamma$ is the fractional loss of PBHs in the cluster, for which the energy is above the escape velocity in the Maxwellian distribution, per relaxation time $t_{\rm relx}$. The evaporation time is roughly the $e$-folding timescale of number reduction according to ${\rm d}N_{\rm cl}/{\rm d}t=-N_{\rm cl}/\Gamma\,t_{\rm relx}=-N_{\rm cl}/t_{\rm evap}$, and $\Gamma\approx135$ for an isolated cluster. Dark matter spikes\;\cite{Adamek:2019gns} or halos with baryon infall\;\cite{Dayal:2025aiv} after $\sim8{\rm\,Myr}~(z\sim160)$ around PBH clusters will increase the escape velocity such that $\Gamma\sim40000$ if the gravitational potential of the bound mass is comparable to the cluster; see Appendix~S.6. In this case, the evaporation time becomes $\sim4{\rm\,Gyr}$, which implies a number reduction of only $1/e^{1/4}\approx0.78$ after $\sim1{\rm\,Gyr}~(z\sim6)$.

\section{Spitzer mass-segregation instability and evaporation time}
To verify the condition of virial equilibrium among different PBH mass populations during evolution, we review the Spitzer mass-segregation instability in a simplified two-component spherical system. Suppose that the equipartition holds
\begin{equation}
\frac{1}{2}m_a\langle v^2\rangle_a=\frac{1}{2}m_b\langle v^2\rangle_b\,,
\end{equation}
with $m_i$ the individual particle mass and $\langle v^2\rangle_i$ the corresponding mean-square velocity, in the subsystem $i\in\{a,b\}$, respectively.
If each subsystem can reach virial equilibrium $2K_i+U_i=0$, then
\begin{equation}
M_i\langle v^2\rangle_i+\frac{1}{2}\int{\rm d}^3\mathbf{x}\,\rho_i(\mathbf{x})\,\phi(\mathbf{x})=0
\quad
\Longrightarrow
\quad
\begin{cases}
\displaystyle
M_a\langle v^2\rangle_a-k_a\frac{GM_a^2}{R_a}
-4\pi\int_0^\infty\frac{GM_b(r)}{r}\rho_a(r)\,r^2{\rm\,d}r=0\,,\\[4pt]
\displaystyle
M_b\langle v^2\rangle_b-k_b\frac{GM_b^2}{R_b}
-4\pi\int_0^\infty\frac{GM_a(r)}{r}\rho_b(r)\,r^2{\rm\,d}r=0\,,
\end{cases}
\end{equation}
where we have used\;\cite{Binney:1987gady}
\begin{equation}
K_i=\frac{1}{2}M_i\langle v^2\rangle_i
\quad{\rm and}\quad
U_i=\frac{1}{2}\int{\rm d}^3\mathbf{x}\,\rho_i(\mathbf{x})\,\phi(\mathbf{x})
=-\int{\rm d}^3\mathbf{x}\,\rho_i(\mathbf{x})\,\mathbf{x}\cdot\nabla\phi(\mathbf{x})
=-4\pi\int_0^R \frac{GM(r)}{r}\rho_i(r)\,r^2{\rm\,d}r\,,
\end{equation}
with the gravitational potential $\phi(\mathbf{x})$ determined by the total mass $M(r)$ mass \emph{both} inside and outside $r$ of the system,
\begin{equation}
\phi(\mathbf{x})=\phi(r)=-\frac{G}{r}\int_0^r{\rm d}M(r')-G\int_r^\infty\frac{{\rm d}M(r')}{r'}\,,
\end{equation}
and $M_i=N_im_i$ is the enclosed mass in radius $R_i$ with $N_i$ the corresponding number of particles, and $k_i=(3-\alpha_i)/(5-2\alpha_i)$ is the structure coefficient if we assume power-law density profiles, $\rho_i(r)=\rho_{0,i}(r/R_i)^{-\alpha_i}$. 

Given that $M_b\ll M_a$ (background system) and $R_b\leq R_a$; while $m_b>m_a$, the above virial equations reduce to
\begin{equation}
\langle v^2\rangle_a=k_a\frac{GM_a}{R_a}
\quad
{\rm and}
\quad
\langle v^2\rangle_b=k_b\frac{GM_b}{R_b}
+\Sigma_{ab}(\alpha_a,\alpha_b)G\rho_{0,a}R_a^{\alpha_a}R_b^{2-\alpha_a}
\end{equation}
with $\Sigma_{ab}(\alpha_a,\alpha_b)=\frac{4\pi (3-\alpha_b)}{(3-\alpha_a)(5-\alpha_a-\alpha_b)}$.
Applying the equipartition condition, $m_a\langle v^2\rangle_a=m_b\langle v^2\rangle_b$, we obtain
\begin{equation}
\frac{m_a}{m_b}k_a\frac{GM_a}{R_a}=k_b\frac{GM_b}{R_b}
+\Sigma_{ab}(\alpha_a,\alpha_b)\,G\,\rho_{0,a}\,R_a^{\alpha_a}\,R_b^{2-\alpha_a}\;.
\end{equation}
For $\alpha_a<2$, the right side, considered as a function of $R_b$, has a minimum value such that
\begin{equation}
\mu\equiv\frac{M_b}{M_a}\leq
\mu_{\rm crit}\equiv\Theta_{ab}(\alpha_a,\alpha_b)\left(\frac{m_a}{m_b}\right)^{\frac{3-\alpha_a}{2-\alpha_a}}
\end{equation}
with
$\Theta_{ab}(\alpha_a,\alpha_b)=\left[\frac{1}{(3-\alpha_b)(5-2\alpha_a)}\right]^{\frac{3-\alpha_a}{2-\alpha_a}}(2-\alpha_a)(5-2\alpha_b)\left(5-\alpha_a-\alpha_b\right)^{\frac{1}{2-\alpha_a}}$.
Note that $\mu_{\rm crit}\propto\left(m_a/m_b\right)^{3/2}$ is the familiar Spitzer scaling when $\alpha_a=0$ (core-like). If this inequality is violated, the virial equilibrium of $b$-subsystem cannot be maintained with $a$-subsystem, and mass segregation proceeds further such that it will eventually decouple from the $a$-subsystem and self-virialize into a dense core. For example, consider $\alpha_a=\alpha_b=0$ for an initial cluster containing $N_{\rm cl}=10^5$ PBHs with $M_a\approx N_{\rm cl}m_{\rm pbh}$. For $b$-subsystem to maintain virial equilibrium, $M_b=N_2 m_{2}=2N_2m_{\rm pbh}$ requires the number of heavy PBHs to remain below $N_{2}\sim6.8\times10^3$; while $M_b=N_{10} m_{10}=10N_{10}m_{\rm pbh}$ gives $N_{10}\lesssim1.2\times10^2$. No PBHs with $m_i\gtrsim m_{60}=60\,m_{\rm pbh}$ can maintain virial equilibrium with the background.

We can also examine how the escape velocity is modified in a two-component system. The escape velocity at $r$ depends on the mass \emph{both} inside and outside $r$ because
\begin{equation}
v_{\rm esc}^2(r)\equiv-2\phi(r)\;.
\end{equation}
When $b$-subsystem is decoupled from the other, it self-virializes and follows its own dispersion with $\langle v^2\rangle_b=k_b GM_b/R_b$. While gravitational potential $\phi(r)$ is determined by the total mass, so that the average escape velocity for the $b$-subsystem is given by
\begin{equation}
\langle v_{\rm esc}^2\rangle_b
=\frac{\int_0^{R_b} v_{\rm esc}^2(r)\rho_b(r)\,r^2{\rm\,d}r}{\int_0^{R_b} \rho_b(r)\,r^2{\rm\,d}r}
=4k_b\frac{GM_b}{R_b}
+\frac{2GM_a/R_a}{2-\alpha_a}
\left[3-\alpha_a-
\frac{3-\alpha_b}{5-\alpha_b-\alpha_a}
\left(\frac{R_b}{R_a}\right)^{2-\alpha_a}\right]\;.
\end{equation}
Therefore, we arrive at
\begin{equation}
\langle v_{\rm esc}^2\rangle_b
=4\langle v^2\rangle_b\left\{1+\frac{(5-2\alpha_b)M_a/R_a}{2(3-\alpha_b)(2-\alpha_a)M_b/R_b}
\left[3-\alpha_a-
\frac{3-\alpha_b}{5-\alpha_b-\alpha_a}
\left(\frac{R_b}{R_a}\right)^{2-\alpha_a}\right]\right\}\;.
\end{equation}
For both core-like distributions, $\alpha_b=\alpha_a=0$, we have
\begin{equation}
\langle v_{\rm esc}^2\rangle_b
=4\langle v^2\rangle_b\left\{1+\frac{5}{4}\frac{M_a/R_a}{M_b/R_b}\left[1-\frac{1}{5}\left(\frac{R_b}{R_a}\right)^2\right]\right\}\;.
\end{equation}
When $M_a=0$, this reduces to the familiar $\langle v_{\rm esc}^2\rangle_b=4\langle v^2\rangle_b$, corresponding to the $0.74\%$ in the Maxwellian tail, thus the fractional loss $1/\Gamma\approx0.74\%$ gives $\Gamma\approx135$. If $M_a/R_a\approx M_b/R_b$ with $R_a\gg R_b$, we have $\langle v_{\rm esc}^2\rangle_b\approx8\langle v^2\rangle_b~(1/\Gamma\approx0.0025\%)$ leading to $\Gamma\sim40000$. This implies that when the gravitational potentials of the $a$- and $b$-subsystems are comparable, the evaporation timescale increases by a factor of $\sim300$.

\section{Orbital dynamics of PBH binaries}\label{Sec:orbit_dyna}
The orbital evolution of a bound binary PBH pair $(i,j)$ is characterized by their impact parameter and initial relative velocity, and here we explicitly use the subscript $ij$ to denote the quantities of the PBH pair in a cluster.

The periapsis, at which the relative acceleration of PBHs is maximized, can be expressed in terms of the relative velocity and PBH masses:
\begin{align}
r_{{\rm p},ij}=\texttt{a}_0(1-\texttt{e}_0)\leq \left(r_{{\rm p},ij}\right)_{\rm max}=\frac{1}{c^2}\left(\frac{85\pi}{6\sqrt{2}}\frac{G^{7/2}\left(m_i+m_j\right)^{3/2}m_im_j}{v_{{\rm rel},ij}^2/c^2}\right)^{2/7}
\end{align}
where $v_{{\rm rel},ij}$ is their relative velocity, $\texttt{a}_0$ and $\texttt{e}_0$ the initial semimajor axis and eccentricity, respectivley. The periapsis is related to the impact parameter $b_{ij}$ by
\begin{align}\label{eq:b_rp}
b_{ij}^2(r_{{\rm p},ij})=\left(r_{{\rm p},ij}\right)^2+\frac{2G(m_i+m_j)r_{{\rm p},ij}}{v_{{\rm rel},ij}^2}\simeq  \frac{2G(m_i+m_j)r_{{\rm p},ij}}{v_{{\rm rel},ij}^2},
\end{align}
where the first term is neglected as $r_{{\rm p},ij}\ll b_{ij}$. Using the above relation, the maximum impact parameter is given by
\begin{align}\label{eq:bmax}
b_{{\rm max},ij}=\left(\frac{2G(m_i+m_j)\left(r_{{\rm p},ij}\right)_{\rm max}}{v_{{\rm rel},ij}^2}\right)^{1/2}
=\left(\frac{85\pi}{6\sqrt{2}}\right)^{1/7}\frac{\sqrt{2}\,G(m_i+m_j)^{5/7}(m_im_j)^{1/7}}{\left(v_{{\rm rel},ij}/c\right)^{9/7}c^2}\,,
\end{align}
and the merging cross section can be written in terms of the impact parameter as
\begin{align}\label{eq:cross_section}
\sigma_{{\rm merg},ij}\equiv\pi b_{{\rm max},ij}^2
=2\pi\left(\frac{85\pi}{6\sqrt{2}}\right)^{2/7}\frac{G^2(m_i+m_j)^{10/7}m_{i}^{2/7}m_{j}^{2/7}}{c^{10/7}v_{{\rm rel},ij}^{18/7}}.
\end{align}

Following sufficient energy loss $\Delta E_{ij}$ due to GW emission, the binary forms a bound state, with final energy:  
\begin{align}
E_{{\rm b},ij}=\frac{1}{2}\mu_{ij}v_{{\rm rel},ij}^2-\Delta E_{ij}
=-\frac{1}{2}\mu_{ij}v_{{\rm rel},ij}^2\left[\left(\frac{b_{{\rm max},ij}}{b_{ij}}\right)^7-1\right]\;.
\end{align}
where $\mu_{ij}\equiv m_im_j/(m_i+m_j)$ is the reduced mass, and we have used
\begin{align}
\Delta E_{ij}=\frac{85\pi}{12\sqrt{2}}\frac{G^{7/2}(m_i+m_j)^{1/2}(m_im_j)^2}{\left(r_{{\rm p},ij}\right)^{7/2}c^5}
=\frac{1}{2}\mu_{ij}v_{{\rm rel},ij}^2\left(\frac{b_{{\rm max},ij}}{b_{ij}}\right)^7.
\end{align}
after applying Eqs.\,\ref{eq:b_rp} and \ref{eq:bmax}.

The initial semimajor axis and eccentricity of the binary can be expressed using the orbital angular momentum $\mu_{ij}\,v_{{\rm rel},ij}\,b_{ij}$ and the binding energy $E_{{\rm b},ij}$:
\begin{align}
\label{eq:a0e0}
&\texttt{a}_0=\frac{G(m_i+m_j)\mu_{ij}}{2|E_{{\rm b},ij}|}=\frac{G(m_i+m_j)}{v_{{\rm rel},ij}^2\left[\left(b_{{\rm max},ij}/b_{ij}\right)^7-1\right]}\,;\notag\\ 
&\texttt{e}_0=\sqrt{1+\frac{2E_{{\rm b},ij}v_{{\rm rel},ij}^2b_{ij}^2}{G^2(m_i+m_j)^2\mu_{ij}}}
=\sqrt{1-\frac{v_{{\rm rel},ij}^4b_{ij}^2\left[\left(b_{{\rm max},ij}/b_{ij}\right)^7-1\right]}{G^2(m_i+m_j)^2}}\;.
\end{align}

After forming a bound state, the binary evolution is characterized by the semimajor axis $\texttt{a}(t)$ and eccentricity $\texttt{e}(t)$, according to the following equations\;\cite{Maggiore:2007ulw}:
\begin{align}\label{eq:eevo_e_a}
\frac{{\rm d}\texttt{a}}{{\rm d}t} &= -\frac{64}{5} \frac{G^3 \mu_{ij} (m_i + m_j)^2}{c^5 \texttt{a}^3} \frac{1}{(1 - \texttt{e}^2)^{7/2}} \left(1 + \frac{73}{24}\texttt{e}^2 + \frac{37}{96}\texttt{e}^4 \right), \notag \\
\frac{{\rm d}\texttt{e}}{{\rm d}t} &= -\frac{304}{15} \frac{G^3 \mu_{ij} (m_i + m_j)^2}{c^5 \texttt{a}^4} \frac{\texttt{e}}{(1 - \texttt{e}^2)^{5/2}} \left(1 + \frac{121}{304}\texttt{e}^2 \right)\;.
\end{align}
Combining two equations in Eq.\,\ref{eq:eevo_e_a}, we obtain
\begin{align}
\frac{{\rm d}\texttt{a}}{{\rm d}\texttt{e}}=\frac{12}{19}\texttt{a}\,\frac{1+\left(\frac{73}{24}\right)\texttt{e}^2+\left(\frac{37}{96}\right)\texttt{e}^4}{\texttt{e}(1-\texttt{e}^2)\left[1+\frac{121}{304}\texttt{e}^2\right]}
\end{align}
which is independent of the mass of the binary. The equation can be solved analytically, and we can obtain the semimajor axis $\texttt{a}$ as a function of the eccentricity $\texttt{e}$ and initial value $\texttt{a}_0$ and $\texttt{e}_0$:
\begin{align}\label{a(e)}
\texttt{a}(\texttt{e},\texttt{e}_0,\texttt{a}_0)=\texttt{a}_0\frac{g(\texttt{e})}{g(\texttt{e}_0)}
\quad{\rm with}\quad
g(\texttt{e})=\frac{\texttt{e}^{\frac{12}{19}}}{1-\texttt{e}^2}\left(1+\frac{121}{304}\texttt{e}^2\right)^{\frac{870}{2299}}.
\end{align}
This in turn determines the evolution of the GW frequency $f_{ij}$ , which is twice the orbital frequency $f_{ij}^{(s)}$ given by Kepler's third law $2\pi f_{ij}^{(s)}=\sqrt{G(m_i+m_j)/\texttt{a}^3}$, and then the minimum GW frequency of the inspiral phase,
\begin{align}
f_{{\rm min},ij}=2f^{(s)}_{{\rm min},ij}=\frac{1}{\pi}\sqrt{\frac{G(m_i+m_j)}{\texttt{a}_0^3}}\;,
\end{align}
where the superscript ``${(s)}$'' denotes the orbital  quantities.

Using Eq.\,\ref{eq:eevo_e_a}, the coalescense time $t_{\rm gw}$ can be obtained by requiring $\texttt{e}(t_{\rm gw})=0$, 
\begin{align}\label{time_coal_e}
\int^{t_{\rm gw}(\texttt{e}_0,\texttt{a}_0)}_0{\rm d}t&=-\frac{15}{304}\frac{c^5}{G^3\mu_{ij}(m_i+m_j)^2}\int_{\texttt{e}_0}^0{\rm d}\texttt{e}\frac{\texttt{a}^4(\texttt{e})\left(1-\texttt{e}^2\right)^{5/2}}{\texttt{e}\left(1 + \frac{121}{304}\texttt{e}^2 \right)}\;.
\end{align}
We then obtain the coalescence time of a binary with $\texttt{e}_0\neq0$:
\begin{align}
t_{\rm gw}(\texttt{e}_0,\texttt{a}_0)
=t_{\rm gw}(\texttt{e}_0=0,\texttt{a}_0)\,F(\texttt{e}_0)
\quad{\rm with}\quad
t_{\rm gw}(\texttt{e}_0=0,\texttt{a}_0)=\frac{5\,c^5\texttt{a}_0^4}{256\,G^3(m_i+m_j)^2\mu_{ij}}\,,
\end{align}
and the enhancement factor
\begin{align}
F(\texttt{e}_0)=\frac{48}{19}\frac{1}{g^4(\texttt{e}_0)}\int^{\texttt{e}_0}_0{\rm d}\texttt{e}\,\frac{g^4(\texttt{e})\left(1-\texttt{e}^2\right)^{5/2}}{\texttt{e}\left(1 + \frac{121}{304}\texttt{e}^2 \right)}
\end{align}
greatly reduces the binary coalescence time when the eccentricity is large. For example, $F(0.9)\simeq 3.4\times10^{-3}$, $F(0.99)\simeq 1.7\times10^{-6}$ and $F(0.999)\simeq 6.1\times10^{-10}$.

\begin{figure}[htbp]
    \centering
    \begin{minipage}{0.49\linewidth}
    \centering
    \includegraphics[width=0.9\linewidth]{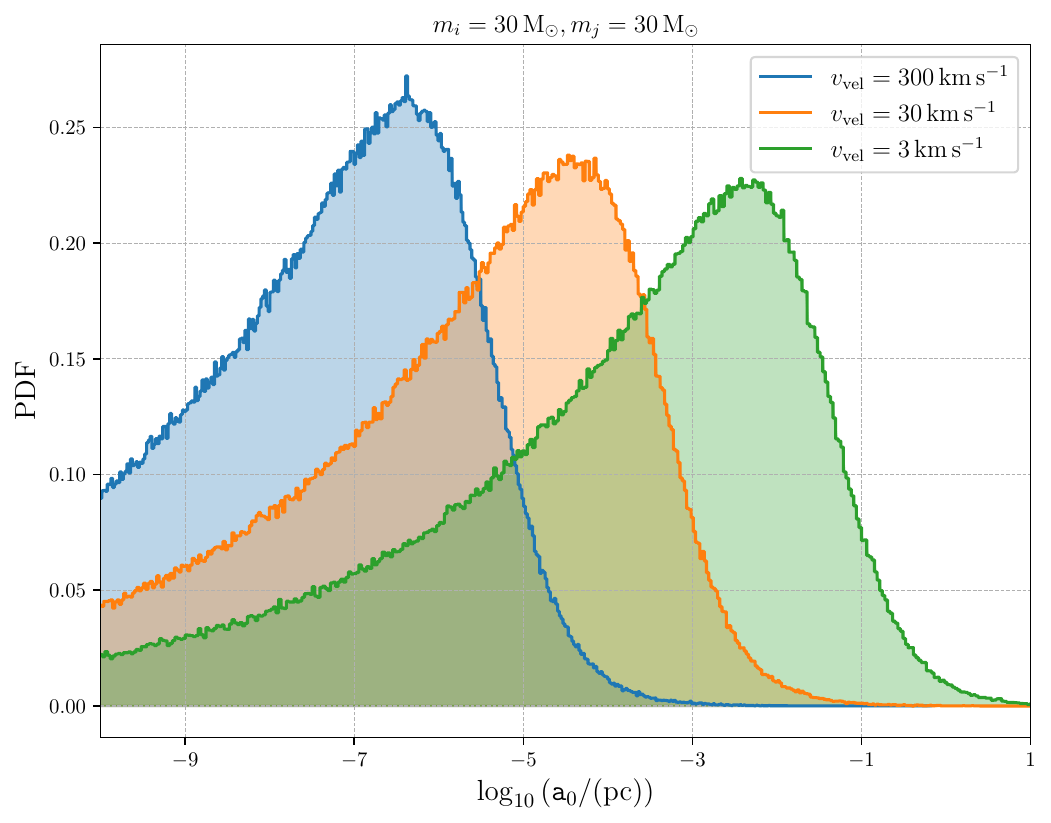}
    \label{PDFe0} 
    \end{minipage}
    \begin{minipage}{0.49\linewidth}
    \centering
    \includegraphics[width=0.9\linewidth]{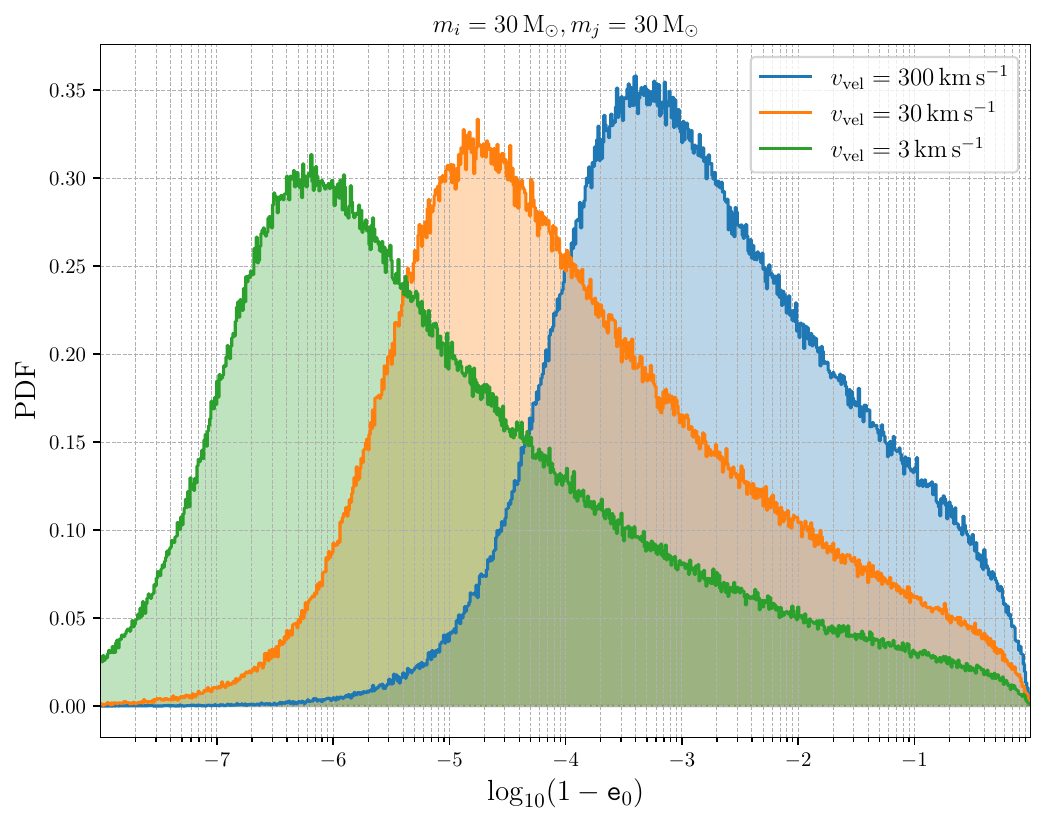}
    \label{PDFa0} 
    \end{minipage}
    \begin{minipage}{0.49\linewidth}
    \centering
    \includegraphics[width=0.9\linewidth]{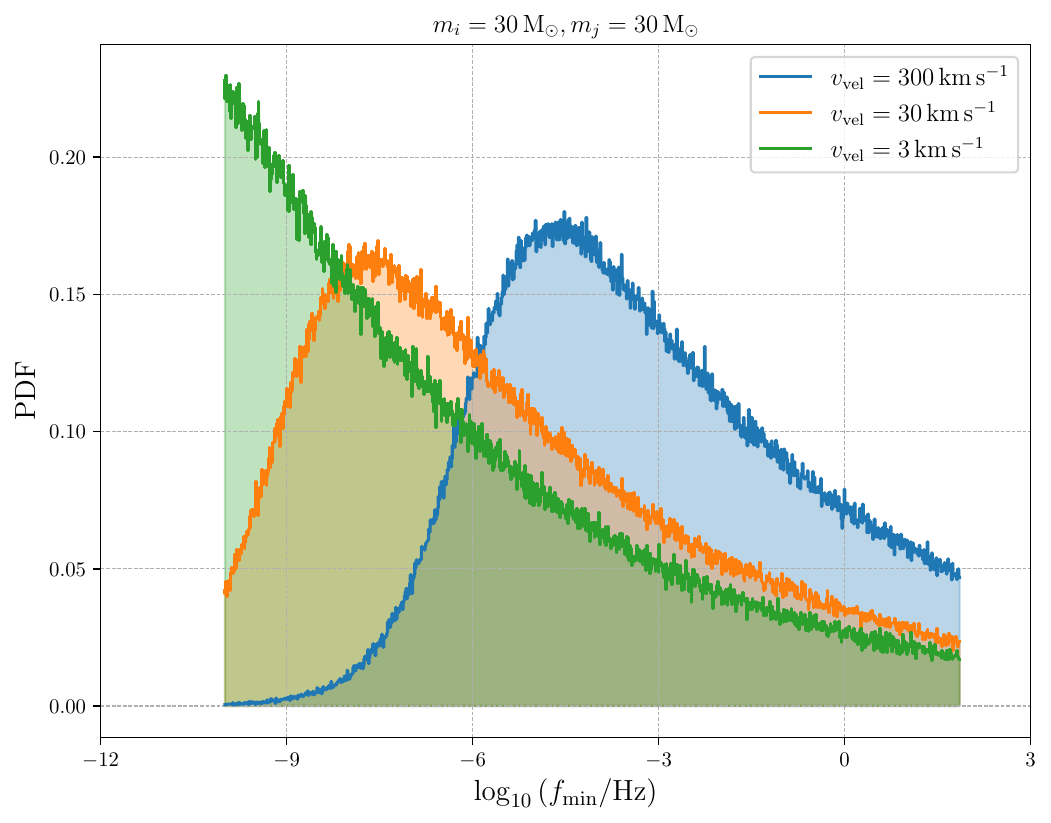}
    \label{PDFfmin} 
    \end{minipage}
    \begin{minipage}{0.49\linewidth}
    \centering
    \includegraphics[width=0.9\linewidth]{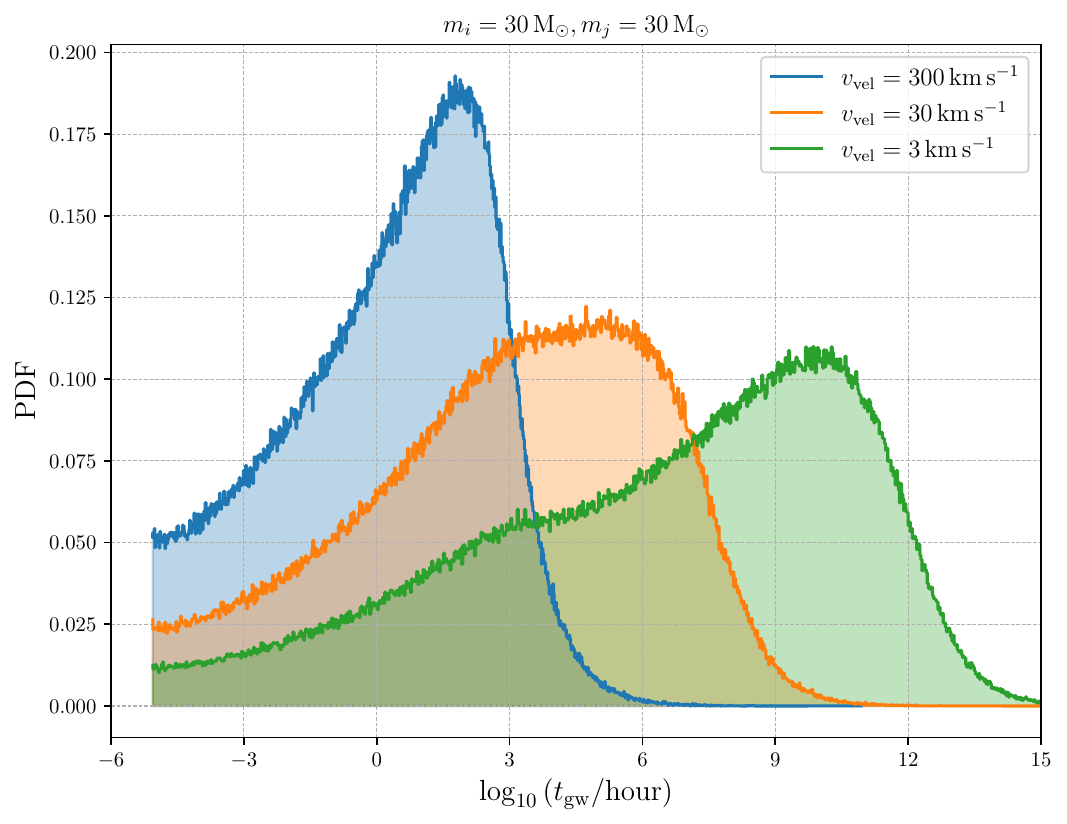}
    \label{PDFtau} 
    \end{minipage}
	\caption{Probability distribution functions (PDFs) of semimajor axis $\texttt{a}_0$, eccentricity $\texttt{e}_0$, minimum frequency $f_{\rm min}$, and coalescence timescale $t_{\rm gw}$.}
	\label{fig:dis_e}
 \end{figure}

We assume that the impact parameter follows a uniform distribution within the range \( b_{ij} \in [0, b_{{\rm max},ij}] \); while the relative velocity follows a Maxwellian distribution with the probability density given by\;\cite{Mouri:2002mc,Zhang:2026vjk}:
\begin{align}\label{eq:velocity_dis}
P(v_{{\rm rel},ij},\alpha_{ij})=\sqrt{\frac{2\alpha_{ij}^3}{\pi}}v_{{\rm rel},ij}^2\exp\left(-\frac{\alpha_{ij} v_{{\rm rel},ij}^2}{2}\right)
\quad{\rm with}\quad\int_0^\infty P(v_{{\rm rel},ij},\alpha_{ij}){\rm~d}v_{{\rm rel},ij}=1
\end{align}
where $\alpha_{ij}=3\mu_{ij}/m_0v_0^2$.
By sampling these orbital parameters, we can derive the eccentricity distribution for each merging PBH pair. Fig.\,\ref{fig:dis_e} shows \(10^6\) binary sample distributions of initial parameters  \(\texttt{a}_0\), \(\texttt{e}_0\), \(f_{\rm min}\), and \(t_{\rm gw}\) given $v_{\rm rel}$ with $m_i=m_j=30{\rm\,M}_\odot$.

\section{Detailed calculations of the GWs from mergers in PBH clusters}
In Ref.\,\cite{Zhang:2026vjk}, we have detailed a Monte Carlo simulation scheme to evolve PBH mergers in a cluster. Here we present how we calculate the integrated stochastic GW background from the PBH merger evolution in clusters.

We follow the standard framework in Ref.\,\cite{Phinney:2001di} to calculate the stochastic GW background. The total present-day energy density of GWs within the frequency interval, ${\rm d}\ln f$, is given by  
\begin{align}
{\rm d}\rho_{\rm gw}(f)\,c^2=\Omega_{\rm gw}(f)\,\rho_cc^2{\rm\,d}\ln f,
\end{align}
with $\Omega_{\rm gw}(f)$ the present-day energy density per logarithmic frequency interval and $\rho_c=\rho_{\rm crit,0}$ the current critical density, and we integrate the contributions of GWs produced by all PBH mergers within this frequency interval across all redshifts:
\begin{align}
{\rm d}\rho_{\rm gw}(f)\,c^2=\left(\int^{\infty}_0\frac{1}{1+z}\mathcal{E}_{\rm gw}(z,f){\rm\,d}z\right){\rm\,d}\ln f
\quad{\rm with}\quad
\mathcal{E}_{\rm gw}(z,f)\equiv \sum_\alpha n(\alpha,z)\frac{{\rm d}E_{\rm gw,\alpha}}{{\rm d}\ln f_r}\Bigg|_{f_r=f(1+z)}
\end{align}
where $1/(1+z)$ is the redshift factor due to comic expansion, $f_r$ denotes the rest-frame GW frequency, $n(\alpha,z)$ is the comoving number density of sourece type $\alpha$ associated with gravitational energy $E_{\rm gw,\alpha}$, and 
thus $\mathcal{E}_{\rm gw}(z,f){\rm\,d}z$ represents the GW energy emitted from sources per unit comoving volume that occurs between redshifts $z$ and $z+{\rm d}z$ contributing to the frequency interval ${\rm d}\ln f$. In the case of PBH cluster, it can be computed as  
\begin{align}
\mathcal{E}_{\rm gw}(z,f){\rm\,d}z&=\bar{n}_{\rm cl}^c\int^{m_{\rm max}(z)}_{m_{\rm min}(z)}\!\!\!{\rm d}m_i\int^{m_{\rm max}(z)}_{m_{\rm min}(z)}\!\!\!{\rm d}m_j\,\langle\sigma_{\rm merg}v_{\rm rel}\rangle\bigg|_{m_i,m_j} \frac{\partial n\left[m_i,t_r(z)\right]}{\partial m_i}\,\frac{\partial n\left[m_j,t_r(z)\right]}{\partial m_j}\,V\frac{{\rm d}t_r}{{\rm d}z}{\rm d}z\notag \\
&\times \left(\frac{{\rm d}E_{{\rm gw},ij}}{{\rm d}\ln f_r}\right)\Bigg|_{f_r=f(1+z)}\theta\left[\ln\left(\frac{f_{{\rm max},ij}}{1+z}\right)-\ln f\right]\,\theta\left[\ln f-\ln\left(\frac{f_{{\rm min},ij}}{1+z}\right)\right],
\end{align}
where $\bar{n}_{\rm cl}^c$ is the average comoving number density of PBH clusters, $n\left(m_i,t_r(z)\right)$ denotes the physical number density of PBHs with mass $m_i$ at time $t_r$, $\mathbf{\rm{\theta}}$ is the Heaviside theta function, $V=N_{\rm cl}/n_{\rm cl}$ is the volume of a cluster, and 
\begin{align}
\frac{{\rm d}t_r}{{\rm d}z}=\frac{1}{H_0(1+z)\sqrt{\Omega_{\rm R}(1+z)^4+\Omega_{\rm M}(1+z)^3+\Omega_\Lambda}}\;.
\end{align}
The merger kernel is given by
\begin{align}
\langle\sigma_{\rm merg}v_{\rm rel}\rangle\bigg|_{m_i,m_j}\equiv\mathcal{K}_{ij}=\mathcal{A}\frac{G^2m_0^2}{c^3}\left(\frac{v_0}{c}\right)^{-11/7}\left(\frac{m_im_j}{m_0^2}\right)^{15/14}\left(\frac{m_i+m_j}{m_0}\right)^{9/14}
\end{align}
with $\mathcal{A}=85^{2/7}(2\pi)^{11/14}3^{1/2}\Gamma(5/7)$.
In our case, the above expression should be discretized, yielding  
\begin{align}
\mathcal{E}_{\rm gw}(z,f){\rm\,d}z=\bar{n}_{\rm cl}^c\sum_{ij} N_{ij}(z){\rm\,d}z\left(\frac{{\rm d}E_{{\rm gw},ij}}{{\rm d}\ln f_r}\right)\bigg|_{f_r=(1+z)f}\theta\left[\ln\left(\frac{f_{{\rm max},ij}}{1+z}\right)-\ln f\right]\,\theta\left[\ln f-\ln\left(\frac{f_{{\rm min},ij}}{1+z}\right)\right]
\end{align}
with $N_{ij}(z)$ the number of merger events for a given pair $(i,j)$ occurring within a redshift interval ${\rm d}z$ defined by
\begin{align}
N_{ij}(z)\equiv Vn(m_i,t_r)\,n(m_j,t_r)\,\mathcal{K}_{ij}\frac{{\rm d}t_r}{{\rm d}z}=N_{i}(t_r)N_j(t_r)\frac{\mathcal{K}_{ij}}{V}\frac{{\rm d}t_r}{{\rm d}z}
\end{align}
and $N_i$ the total number of PBHs with mass $m_i$ in a cluster. 

In the circular orbit approximation, 
the gravitational radiation emitted by a single PBH binary in the Newtonian limit:
\begin{align}
\frac{{\rm d}E_{{\rm gw},ij}}{{\rm d}\ln f_r}=\frac{\left(G\mathcal{M}_{ij}\right)^{5/3}\left(\pi f_r\right)^{2/3}}{3\,G},\quad f_{{\rm min},ij}<f_r<f_{{\rm max},ij}
\end{align}
where $\mathcal{M}_{ij}=\left(m_im_j\right)^{3/5}/\left(m_i+m_j\right)^{1/5}$ denotes the chirp mass of a PBH pair with masses $m_i$ and $m_j$. The frequency of GWs in the source’s cosmic rest frame, $f_r$, is related to the observed frequency today by $f_r=(1+z)f$. The minimum frequency, $f_{{\rm min},ij}=2f_{{\rm min},ij}^{(s)}(t=0)$, is determined by the initial separation of the binary, which must be small enough for the binary to become gravitationally bound, ensuring that the merger occurs within the age of the Universe. In our case, the high PBH number density within a cluster results in a merger timescale for a PBH pair that is significantly shorter than the overall cluster evolution timescale, which itself remains much smaller than the age of the Universe. $f_{{\rm max},ij}$ is the maximum frequency, identified as twice the revolution frequency in the innermost stable orbit (ISCO)
\begin{align}
f_{\mathrm{max},ij}=2f^{(s)}_{{\rm max},ij}=2f_{\mathrm{ISCO},ij}\simeq \frac{c^3}{6^{3/2}G\pi \left(m_i+m_j\right)}\;.
\end{align}
Thus,
\begin{align}\label{eq:Omegagw}
\Omega_{\rm gw}(f)
&=\frac{1}{\rho_cc^2}\int^{\infty}_0\frac{\mathcal{E}_{\rm gw}(z,f){\rm\,d}z}{1+z}\notag\\
&=\frac{(G\pi f)^{2/3}}{3\,\rho_cc^2}\bar{n}_{\rm cl}^c\int^{\infty}_0 \frac{{\rm d}z}{(1+z)^{1/3}}\sum_{ij}N_{ij}(z)\,\mathcal{M}_{ij}^{5/3}\theta\left[\ln\left(\frac{f_{{\rm max},ij}}{1+z}\right)-\ln f\right]\,\theta\left[\ln f-\ln\left(\frac{f_{{\rm min},ij}}{1+z}\right)\right]\;.
\end{align}
Provided that the sources are randomly oriented with respect to earth, the formula does not depend on properties such as beaming and polarization of the sources. 

Bound through two-body capture, PBH binaries  generally have large eccentricities $\texttt{e}_0\approx1$.  This actually modifies the GW frequency spectrum. For nonzero eccentricity, the GW emission is no longer monochromatic at \(\omega_{\rm gw} = 2\,\omega^{(s)}\); instead, power is distributed across harmonics \(\omega_{\mathrm{gw},n} = n\,\omega^{(s)}\) for integer \(n\). The harmonic number at which the spectrum peaks increases with eccentricity. As a result, compared with circular orbits, power is removed from the \(n=2\) mode and transferred to higher frequencies.

We compute the stochastic GW background by summing the contributions of all harmonics. For the \(n\)-th harmonic, the emitted energy per logarithmic frequency interval in the source rest frame is given by~\cite{Maggiore:2007ulw,Maggiore:2018sht}:
\begin{align}
\frac{{\rm d}E_{\mathrm{gw},ij,n}}{{\rm d}\ln f_{n,r}}=\frac{{\rm d}E_{\mathrm{gw},ij,n}}{{\rm d}t}\frac{{\rm d}t}{{\rm d}\texttt{a}}\frac{{\rm d}\texttt{a}}{{\rm d}\ln f_r},
\end{align}
Here the \(n\)-th harmonic frequency is \(f_{n,r} = n\,f_r/2\), thus \({\rm d}\ln f_{n,r} = {\rm d}\ln f_r\). The radiated power in the \(n\)-th harmonic is given by
\begin{align}\label{pn}
\frac{{\rm d}E_{\mathrm{gw},ij,n}}{{\rm d}t}=\frac{32G^4\mu_{ij}^2(m_i+m_j)^3}{5c^5\texttt{a}^5}\mathcal{G}(n,\texttt{e})
\end{align}
with the Peters–Mathews spectral function\;\cite{Maggiore:2007ulw}
\begin{align}\label{eq:G(n,e)}
\mathcal{G}(n,\texttt{e})&=\frac{n^6}{96\,\texttt{a}^4}\bigl[A_n^2(\texttt{e})+B_n^2(\texttt{e})+3C^2_n(\texttt{e})-A_n(\texttt{e})B_n(\texttt{e})\bigr],\notag\\
A_n(\texttt{e})&=\frac{\texttt{a}^2}{n}\Bigl\{J_{n-2}(n\texttt{e})-J_{n+2}(n\texttt{e})-2\texttt{e}\bigl[J_{n-1}(n\texttt{e})-J_{n+1}(n\texttt{e})\bigr]\Bigr\},\notag\\
B_n(\texttt{e})&=\frac{\texttt{b}^2}{n}\bigl[J_{n+2}(n\texttt{e})-J_{n-2}(n\texttt{e})\bigr],\notag\\
C_n(\texttt{e})&=\frac{\texttt{ab}}{n}\Bigl\{J_{n+2}(n\texttt{e})+J_{n-2}(n\texttt{e})-\texttt{e}\bigl[J_{n+1}(n\texttt{e})+J_{n-1}(n\texttt{e})\bigr]\Bigr\},
\end{align}
where $J_{\alpha}(x)$ is Bessel functions of the first kind and the semiminor axis $\texttt{b}=\texttt{a}\sqrt{1-\texttt{e}^2}$. In the case of circular orbits, only the second harmonic contributes, $\mathcal{G}(n,0)=\delta_{2n}$.
Using Kepler's law, Eq.\,\ref{eq:eevo_e_a} and Eq.\,\ref{pn} , we obtain 
\begin{align}
\frac{{\rm d}E_{\mathrm{gw},ij,n}}{{\rm d}\ln f_{n,r}}=\frac{\left(G\mathcal{M}_{ij}\right)^{5/3}}{3\,G}\left(\frac{2\pi f_{n,r}}{n}\right)^{2/3}\frac{\mathcal{G}(n,\texttt{e})}{\mathcal{F}(\texttt{e})}
\quad{\rm with}\quad
\mathcal{F}(\texttt{e})=\frac{1}{\left(1-\texttt{e}^2\right)^{7/2}}\left(1+\frac{73}{24}\texttt{e}^2+\frac{37}{96}\texttt{e}^4\right)\;.
\label{eq:F(e)}
\end{align}

To construct the stochastic background, we partition the observed frequency range into bins. Within each bin, we sum the GW contributions from all harmonic modes of binaries across every redshift slice for which the emitted harmonic frequency falls within that bin. Consequently,
\begin{align}
\mathcal{E}_{\rm gw}(z,f){\rm\,d}z
&=\bar{n}_{\rm cl}^c\sum_{ij} N_{ij}(z){\rm\,d}z\sum_{n=1}^{\infty}\frac{{\rm d}E_{\mathrm{gw},ij,n}}{{\rm d}\ln f_{n,r}}\bigg|_{f_{n,r}=\frac{n}{2}(1+z)f}\theta\left(\ln f^{(s)}_{\mathrm{max},ij}-\ln f^{(s)}_{n}\right)\theta\left(\ln f^{(s)}_{n}-\ln f^{(s)}_{\mathrm{min},ij}\right)\notag\\
&=\bar{n}_{\rm cl}^c\sum_{ij} N_{ij}(z){\rm\,d}z\sum_{n=1}^{\infty}\frac{{\rm d}E_{\mathrm{gw},ij,n}}{{\rm d}\ln f_{n,r}}\bigg|_{f_{n,r}=\frac{n}{2}(1+z)f}\theta\left[\ln\left(\frac{nf_{{\rm max},ij}}{2(1+z)}\right)-\ln f\right]\theta\left[\ln f-\ln\left(\frac{nf_{{\rm min},ij}}{2(1+z)}\right)\right]\notag\\
&=\bar{n}_{\rm cl}^c\sum_{ij} N_{ij}(z){\rm\,d}z\frac{\left(G\pi f\right)^{2/3}\mathcal{M}_{ij}^{5/3}}{3}(1+z)^{2/3}\,\Upsilon(\texttt{e},f,z,m_i,m_j)
\end{align}
with 
\begin{align}\label{Phie}
\Upsilon(\texttt{e},f,z,m_i,m_j)=\sum_{n=1}^{\infty}\left(\frac{n}{2}\right)^{-2/3}\frac{\mathcal{G}(n,\texttt{e})}{\mathcal{F}(\texttt{e})}\theta\left[\ln\left(\frac{nf_{{\rm max},ij}}{2(1+z)}\right)-\ln f\right]\,\theta\left[\ln f-\ln\left(\frac{nf_{{\rm min},ij}}{2(1+z)}\right)\right]\;.
\end{align}
Here, the observed frequency $f$ is related by $f=f_r/(1+z)=nf^{(s)}_n/(1+z)$, where \(f^{(s)}_n\) denotes the orbital frequency corresponding to the \(n\)-th harmonic at rest-frame frequency \(f_r\). Other than the familiar \(f^{2/3}\) scaling of circular orbits, $\texttt{a}(t)$ and $\texttt{e}(t)$ evolving with time will introduce extra frequency-dependence. Using Kepler's law, $\texttt{a}(t)$ can be written in terms of the instantaneous orbital frequency of the source $f^{(s)}(t)$, then the eccentricity is a function of frequency which can be written as 
\begin{align}
\texttt{e}=\texttt{e}(f^{(s)}(t),\texttt{e}_0,\texttt{a}_0)=\texttt{e}\left(\frac{(1+z)f}{n},\texttt{e}_0,\texttt{a}_0\right),
\end{align}
which induces the $f$ dependence thus modifies the GW spectrum. Using Kepler's third law and Eq.\;\ref{a(e)}, we can obtain the relation between $\texttt{e}$ and $f^{(s)}$~\cite{Maggiore:2007ulw}
\begin{align}\label{e(f)}
g(\texttt{e})=g(\texttt{e}_0)\left(\frac{f^{(s)}_{0}}{f^{(s)}}\right)^{2/3}.
\end{align}
By interpolating this relation, we determine \(\texttt{e}(f^{(s)})\) and compute resulting energy density spectrum
\begin{align}\label{eq:e_Omegagw}
\Omega_{\rm gw}(f)&=\frac{1}{\rho_cc^2}\int^{\infty}_0\frac{\mathcal{E}_{\rm gw}(z,f){\rm\,d}z}{1+z}\notag\\
&=\frac{\left(G\pi f\right)^{2/3}}{3\,\rho_cc^2}\bar{n}_{\rm cl}^c\int^{\infty}_0{\rm d}z\,(1+z)^{-1/3}\sum_{ij}\mathcal{M}_{ij}^{5/3} N_{ij}(z)\,\Upsilon\left[\texttt{e}\left(\frac{(1+z)f}{n},\texttt{e}_0,\texttt{a}_0\right),f,z,m_i,m_j\right].
\end{align}

The validity of Eqs.\,\ref{eq:Omegagw} ($\texttt{e}=0$) and \ref{eq:e_Omegagw} ($\texttt{e}\neq0$) hinge on the fact that the binary lifetime (or coalescence timescale) is sufficiently short compared to the cosmic expansion timescale, and thus the entire inspiral occurs at the same redshift as two PBHs form a bound state. This approximation is justified, as shown in Fig.\,\ref{fig:dis_e}. For runaway mergers to occur, the cluster must be sufficiently dense, with a virial velocity of order $\sim \mathcal{O}(100){\rm\,km/s}$. Under these conditions, the coalescence timescale is indeed very short, and in some cases less than a year when eccentricity is taken into account. Moreover, this approximation is crucial for ensuring the validity of our Monte Carlo simulations\;\cite{Zhang:2026vjk}. 

\section{Infrared scaling of $\Omega_{\rm gw}$} 
In Eq.\,\ref{eq:e_Omegagw}, the frequency-dependent part beyond the standard $f^{2/3}$ scaling is encapsulated in $\Upsilon(\texttt{e},f,z,m_i,m_j)$. We can examine the infrared scaling of the GW spectrum analytically. In this regime, nearly all sources contribute to the GW spectrum, since their maximum frequencies exceed the upper bound and their minimum frequencies lie below the lower bound of the considered range. Consequently, the Heaviside step function in Eq.\,\ref{Phie} can be omitted, and the factor reduces to
\begin{align}\label{eq:Upsilon}
\Upsilon(\texttt{e},f,z,m_i,m_j)=\sum_{n=1}^{\infty}\left(\frac{n}{2}\right)^{-2/3}\frac{\mathcal{G}(n,\texttt{e})}{\mathcal{F}(\texttt{e})}.
\end{align}
We compute the derivative of this factor with respect to \(f\), which yields
\begin{align}
\frac{{\rm d}\Upsilon}{{\rm d} f}=\sum_{n=1}^{\infty}\left(\frac{n}{2}\right)^{-2/3}\frac{\mathcal{G}(n,\texttt{e})}{\mathcal{F}(\texttt{e})}\left(\frac{1}{\mathcal{G}(n,\texttt{e})}\frac{{\rm d}\mathcal{G}(n,\texttt{e})}{{\rm d}\texttt{e}}-\frac{1}{\mathcal{F}(\texttt{e})}\frac{d\mathcal{F}(\texttt{e})}{d\texttt{e}}\right)\frac{{\rm d}\texttt{e}}{{\rm d}f}.
\end{align}
Using \(\mathcal{G}(n,\texttt{e})\) in Eq.\,\ref{eq:G(n,e)} and the formula,
\begin{align}
\frac{{\rm d}J_{q}(x)}{{\rm d}x}=\frac{1}{2}\left(J_{q-1}(x)-J_{q+1}(x)\right)\,,
\end{align}
we can compute
\begin{align}
\frac{{\rm d}\mathcal{G}(n,\texttt{e})}{{\rm d}\texttt{e}}=\frac{n^6}{96\,\texttt{a}^4}\frac{1}{n^2}\left(\texttt{a}^4\text{-term}_{d}+\texttt{b}^4\text{-term}_{d}+\texttt{a}^2\texttt{b}^2\text{-term}_{d}\right),
\end{align}
where the individual terms are
\begin{align}
\texttt{a}^4\text{-term}_d
&=-\frac{32\,\texttt{a}^4}{\texttt{e}^5 n^2}
\Bigl[\texttt{e}(-1+\texttt{e}^2)n\,J_{n-1}(n\texttt{e})+(1+n-\texttt{e}^2n)\,J_n(n\texttt{e})\Bigr]\notag\\
&\qquad\quad\times
\Bigl\{-3\texttt{e} n\,J_{n-1}(n\texttt{e})+\Bigl[2+n\bigl(3+(-1+\texttt{e}^2)^2 n\bigr)\Bigr]\,J_n(n\texttt{e})\Bigr\},
\\
\texttt{b}^4\text{-term}_d
&=\texttt{b}^4\,n
\bigl[J_{n-2}(n\texttt{e})-J_{n+2}(n\texttt{e})\bigr]\notag\\
&\qquad\quad\times
\bigl[J_{n-3}(n\texttt{e})-J_{n-1}(n\texttt{e})-J_{n+1}(n\texttt{e})+J_{n+3}(n\texttt{e})\bigr],
\\
\texttt{a}^2\texttt{b}^2\text{-term}_d
&=\frac{16\,\texttt{a}^2\texttt{b}^2}{\texttt{e}^5 n^2}
\Bigl[
3\texttt{e}^2\bigl(-4+(-4+3\texttt{e}^2)n^2\bigr)\,J_{n-1}(n\texttt{e})^2\notag\\
&\qquad\;+\;2\texttt{e}\Bigl(12+4\texttt{e}^4n^3+4n(5+n(3+n))-\texttt{e}^2n\bigl(10+n(9+8n)\bigr)\Bigr)\,J_{n-1}(n\texttt{e})J_n(n\texttt{e})\notag\\
&\qquad\;-\;\Bigl(16+n\bigl(40+8n(4+n)-4\texttt{e}^2(1+n)(5+4n)+\texttt{e}^4n(7+8n)\bigr)\Bigr)\,J_n(n\texttt{e})^2\Bigr]\;.
\end{align}
Similarly, we write
\begin{align}
\mathcal{G}(n,\texttt{e})=\frac{n^6}{96\,\texttt{a}^4}\frac{1}{n^2}\left(\texttt{a}^4\text{-term}+\texttt{b}^4\text{-term}+\texttt{a}^2\texttt{b}^2\text{-term}\right),
\end{align}
where the corresponding terms are
\begin{align}
\texttt{a}^4\text{-term}
&=\texttt{a}^4\bigl(J_{n-2}(n\texttt{e})-4\texttt{e}\,J_{n-1}(n\texttt{e})+4\,J_n(n\texttt{e})-J_{n+2}(n\texttt{e})\bigr)^2\,,\\
\texttt{b}^4\text{-term}
&=2\,\texttt{b}^4\bigl(J_{n-2}(n\texttt{e})-J_{n+2}(n\texttt{e})\bigr)^2\,,\\
\texttt{a}^2\texttt{b}^2\text{-term}
&=\texttt{a}^2\texttt{b}^2\Bigl[\,\bigl(J_{n-2}(n\texttt{e})-J_{n+2}(n\texttt{e})\bigr)\,\bigl(J_{n-2}(n\texttt{e})-4\texttt{e}\,J_{n-1}(n\texttt{e})+4\,J_n(n\texttt{e})-J_{n+2}(n\texttt{e})\bigr)\notag\\
&\qquad+\;3\bigl(J_{n-2}(n\texttt{e})-2\,J_n(n\texttt{e})+J_{n+2}(n\texttt{e})\bigr)^2\Bigr]\;. 
\end{align}
From Eq.\,\ref{eq:F(e)}, we obtain
\begin{align}
\frac{1}{\mathcal{F}(\texttt{e})}\frac{{\rm d}\mathcal{F}(\texttt{e})}{{\rm d}\texttt{e}}=\frac{\texttt{e}(1256+1608\texttt{e}^2+111\texttt{e}^4)}{96+196\texttt{e}^2-255\texttt{e}^4-37\texttt{e}^6}. 
\end{align}
The final derivative, \({\rm d}\texttt{e}/{\rm d}f\), follows from Eq.\,\ref{e(f)} and the relation \(f=n\,f^{(s)}/(1+z)\):
\begin{align}
\frac{{\rm d}f}{{\rm d}\texttt{e}}=-\frac{3}{2}\frac{n}{1+z}f^{(s)}_{0}&\left(\frac{g(\texttt{e}_0)}{g(\texttt{e})}\right)^{3/2}\frac{1}{g(\texttt{e})}\frac{{\rm d}g(\texttt{e})}{{\rm d}\texttt{e}}
=-\frac{3}{2}f\frac{192+584\texttt{e}^2+74\texttt{e}^4}{304\texttt{e}-183\texttt{e}^3-121\texttt{e}^5}\notag\\
\quad\Rightarrow\quad
&\frac{{\rm d}\texttt{e}}{{\rm d}f}=-\frac{2}{3}\frac{1}{f}\frac{304\texttt{e}-183\texttt{e}^3-121\texttt{e}^5}{192+584\texttt{e}^2+74\texttt{e}^4}.
\end{align}
Since in the infrared regime eccentricity remains near its initial value, and most binaries are highly eccentric (\(\texttt{e}\approx1\)), we take the limit \(\texttt{e}\to1\), which yields
\begin{align}
\lim_{\texttt{e}\to 1}\frac{1}{\mathcal{G}(n,\texttt{e})}\frac{{\rm d}\mathcal{G}(n,\texttt{e})}{{\rm d}\texttt{e}}=-4+6n\left(\frac{J_{n-1}(n)}{J_n(n)}-1\right)\,,
\quad
\lim_{\texttt{e}\to 1} \frac{{\rm d}\texttt{e}}{{\rm d}f}=0\,,
\quad
\lim_{\texttt{e}\to 1}\frac{1}{\mathcal{F}(\texttt{e})}\frac{{\rm d}\mathcal{F}(\texttt{e})}{{\rm d}\texttt{e}}\frac{{\rm d}\texttt{e}}{{\rm d}f}=-\frac{1}{f}\frac{7}{3}\;.
\end{align}
Although both the denominator of \(\frac{1}{\mathcal{F}(\texttt{e})}\frac{{\rm d}\mathcal{F}(\texttt{e})}{{\rm d}\texttt{e}}\) and the numerator of \(\frac{{\rm d}\texttt{e}}{{\rm d}f}\) vanish as \(\texttt{e}\to1\), they cancel, yielding a finite result. Thus, the result reduces to
\begin{align}
\lim_{\texttt{e}\to1} \left(\frac{1}{\mathcal{G}(n,\texttt{e})}\frac{{\rm d}\mathcal{G}(n,\texttt{e})}{{\rm d}\texttt{e}}-\frac{1}{\mathcal{F}(\texttt{e})}\frac{{\rm d}\mathcal{F}(\texttt{e})}{{\rm d}\texttt{e}}\right)\frac{{\rm d}\texttt{e}}{{\rm d}f}=0-\left(-\frac{1}{f}\frac{7}{3}\right)=\frac{1}{f}\frac{7}{3}.
\end{align}
Consequently, the logarithmic slope becomes
\begin{align}
\frac{{\rm d}\ln\Omega_{\rm gw}}{{\rm d}\ln f}=\frac{2}{3}+\frac{7}{3}=3\,,
\end{align}
which explains the infrared behavior of $\Omega_{\rm gw}\propto f^3$. It is interesting to note that this scaling is the same as the universal infrared scaling in the radiation-dominated background\;\cite{Cai:2019cdl}, but due to different origins.

\end{document}